\documentclass[prb,twocolumn,superscriptaddress,float,aps]{revtex4-2}
\usepackage{graphicx,amsfonts,amssymb,amsmath,hyperref,hypcap,enumerate}
\usepackage{color}	
\usepackage{dsfont}
\usepackage{bm}
\usepackage{multirow}
\usepackage{makecell}
\usepackage[normalem]{ulem}

\begin{document}
\title{Spin-orbit-coupled superconductivity with spin-singlet non-unitary pairing}
 
\author{Meng Zeng}
\affiliation{Department of Physics, University of California, San Diego, California 92093, USA}

\author{Dong-Hui Xu}
\affiliation{Department of Physics and Chongqing Key Laboratory for Strongly Coupled Physics, Chongqing University, Chongqing 400044, China}
\affiliation{Center of Quantum Materials and Devices, Chongqing University, Chongqing 400044, China}

\author{Zi-Ming Wang}
\affiliation{Department of Physics and Chongqing Key Laboratory for Strongly Coupled Physics, Chongqing University, Chongqing 400044, China}
\affiliation{Center of Quantum Materials and Devices, Chongqing University, Chongqing 400044, China}

\author{Lun-Hui Hu}
\thanks{hu.lunhui.zju@gmail.com}
\affiliation{Department of Physics, the Pennsylvania State University, University Park, PA, 16802, USA}
\affiliation{Department of Physics and Astronomy, University of Tennessee, Knoxville, Tennessee 37996, USA}

\begin{abstract}
The gap functions for a single-band model for unconventional superconductivity are distinguished by their unitary or non-unitary forms.
Here we generalize this classification to a two-band superconductor with two nearly degenerate orbitals. We focus on spin-singlet pairings and investigate the effects of the atomic spin-orbit coupling (SOC) on superconductivity which is a driving force behind the discovery of a new spin-orbit-coupled non-unitary superconductor. Multi-orbital effects like orbital hybridization and strain induced anisotropy will also be considered.
The spin-orbit-coupled non-unitary superconductor has three main features.
First, the atomic SOC locks the electron spins to be out-of-plane,
leading to a new Type II Ising superconductor with a large in-plane upper critical field beyond the conventional Pauli limit.
Second, it provides a promising platform to realize the topological chiral or helical Majorana edge state even without external magnetic fields or Zeeman fields.
More surprisingly, a spin-polarized superconducting state could be generated by spin-singlet non-unitary pairings when time-reversal symmetry is spontaneously broken, which serves as a smoking gun to detect this exotic state by measuring the spin-resolved density of states. 
Our work indicates the essential roles of orbital-triplet pairings in both unconventional and topological superconductivity.
\end{abstract}

\maketitle

\section{Introduction}

In condensed matter physics, research on unconventional superconductivity~\cite{sigrist_rmp_1991,mineev1999} remains a crucial topic and continues to uncover new questions and challenges in both theory and experiment, since the discovery of the heavy-fermion superconductors (SCs)~\cite{steglich_prl_1979} and the $d$-wave pairing states in high-temperature cuprate SCs~\cite{Bednorz1986,zhang_prb_1988,anderson_jpcm_2004,scalapino_rmp_2012}.
In addition to the anisotropic gap functions (e.g., $p, d, f, g$-wave...), the sublattice or orbital-dependent pairings~\cite{brydon_prx_2019,jose_arxiv_2021,zeng_arxiv_2021b} are shown to be an alternative avenue to realize unconventional SCs.
They might be realized in multi-orbital correlated electronic systems, whose candidate materials include iron-based SCs \cite{dai_prl_2008,ong_prl_2013,yin_natphy_2014,ong_pnas_2016,nourafkan_prl_2016,chubukov_prb_2016,yi_npjqm_2017,nica_npjqm_2017,sprau_sci_2017,hu_prb_2018,chen_prr_2021,nica_npjqm_2021},
Cu-doped Bi$_2$Se$_3$ \cite{wray_np_2010,fu_prl_2010},
half-Heusler compounds \cite{brydon_prl_2016,yang_prl_2016,agterberg_prl_2017,savary_prb_2017,yang_prb_2017,timm_prb_2017,yu_prb_2018,boettcher_prl_2018,roy_prb_2019,kim_sciadv_2018}, and possibly Sr$_2$RuO$_4$ \cite{agterberg_prl_1997,takimoto_prb_2000,huang_prb_2019,wang_arxiv_xxx,Clepkens1,Ramires_prb_2019} etc.
In particular, considering the atomic orbital degrees of freedom, the classification of unconventional pairing states could be significantly enriched.
Among them, SCs with spontaneous time-reversal symmetry (TRS) breaking is of special interest, in which two mutually exclusive quantum phenomena, spin magnetism, and superconductivity may coexist with each other peacefully\cite{yang2017,yuan_prb_2017,chirolli_prb_2017,robins2018,lado_prr_2019,hu_prr_2020}.

On the other hand, the orbital multiplicity could also give rise to non-unitary pairings, which again include both time-reversal breaking (TRB) and time-reversal invariant (TRI) pairings. Very recently, prior studies have demonstrated the existence of spin-singlet non-unitary pairing states that break the inversion symmetry in Dirac materials~\cite{jose_arxiv_2021}. 
One aim of this work is the generalization of unitary and non-unitary gap functions in a two-band SC while preserving inversion symmetry, which is possible exactly due to the multi-orbital degrees of freedom~\cite{ramires_JPCM_2022}.
We focus on a system with two nearly degenerate  orbitals and find that the non-unitary pairing state is generally a mixed superconducting state with both orbital-independent pairings and orbital-dependent pairings. 
Recently, the interplay between orbital-independent pairings and spin-orbit coupling (SOC) has been shown to demonstrate the intriguing phenomenon of a large in-plane upper critical field compared with the Pauli paramagnetic field for a two-dimensional SC.
For example, the Type I Ising superconductivity in monolayer MoS$_2$ \cite{saito2016superconductivity,lu2015evidence} and NbSe$_2$ \cite{xi2016ising} and the Type II Ising superconductivity in monolayer stanene \cite{falson2020type}.
Therefore, the interplay of atomic SOC and the multi-orbital pairing could potentially give rise to exciting physics. However, to the best of our knowledge, the influence of the atomic SOC on the orbital-dependent pairings remains unsolved. 
Furthermore, the multi-orbital nature also gives rise to possible orbital hybridization effects and provides an experimentally controllable handle using lattice strains, both of which could lead to orbital anisotropy and could potentially change the pairing symmetry. In particular, lattice strain has been a useful experimental tool to study unconventional superconductors \cite{ruf2021strain,beck2022effectsstrain,ahadi2019strainenhancing} and has even been proposed to induce the elusive charge-4e phase \cite{fernandes2021charge4e}. We will be doing an extensive investigation on all the aforementioned multi-orbital effects.

Another topic of this work is concerned with the coexistence of TRB pairings and spin magnetism even in a spin-singlet SC. 
It is well-known that spin-polarization (SP) can be generated by nonunitary \textit{spin-triplet} superconductivity, which is believed to be the case for LaNiC$_2$ \cite{hillier2009} and LaNiGa$_2$ \cite{hillier2012,weng2016}.
More recently, the coexistence of magnetism and \textit{spin-singlet} superconductivity is experimentally suggested in multi-orbital SCs, such as iron-based superconductors \cite{grinenko2020,zaki2019} and LaPt$_3$P \cite{biswas_nc_2021}.
Therefore, in addition to the spin-triplet theory, it will be interesting to examine how SP develops in multi-orbital spin-singlet SCs as spontaneous TRS breaking in the absence of external magnetic fields or Zeeman fields.

In this work, we address the above two major issues by studying a two-band SC with two atomic orbitals (e.g., $d_{xz}$ and $d_{yz}$).
We start with the construction of a ${\bf k}\cdot{\bf p}$ model Hamiltonian on a square lattice with applied lattice strain. The breaking of $C_{4v}$ down to $C_2$ point group generally leads to the degeneracy lifting of $d_{xz}$ and $d_{yz}$. Based on this model, we study the stability of superconductivity and the realization of 2D topological superconductors in both class D and DIII. 
First and foremost, the influence of atomic SOC is studied, which gives birth to a new spin-orbit-coupled SC.
This exotic state shows the following features:
firstly, a large Pauli-limit violation is found for the orbital-independent pairing part, which belongs to the Type II Ising superconductivity.
Furthermore, the orbital-dependent pairing part also shows a weak Pauli-limit violation even though it does not belong to the family of Ising SCs.
Secondly, topological superconductivity can be realized with a physical set of parameters even in the absence of external magnetic fields or Zeeman fields.
In addition, a spin-polarized superconducting state could be energetically favored with the spontaneous breaking of time-reversal symmetry. 
Our work implies a new mechanism for the establishment of spin magnetism in the spin-singlet SC. 
In the end, we also discuss how to detect this effect by spin-resolved scanning tunneling microscopy measurements.

The paper is organized as follows: 
in section~\ref{sec-two}, we discuss a two-orbital normal-state Hamiltonian on a 2D square lattice and also its variants caused by applied in-plane strain effects, then we show the spin-singlet unitary or non-unitary pairing states with or without TRS. The strain effect on pairing symmetries is also studied based on a weak-coupling theory.
In section~\ref{sec-three}, the effects of atomic SOC on such pairing states are extensively studied, as well as the in-plane paramagnetic depairing effect.
Besides, the topological superconductivity is studied in section~\ref{sec-four} even in the absence of external magnetic fields or Zeeman fields, after which we consider the spontaneous TRB effects in section~\ref{sec-five} and show that spin-singlet SC-induced spin magnetism could emerge in the presence of orbital SOC.
In the end, a brief discussion and conclusion are given in section~\ref{sec-six}.
We will also briefly comment on a very recent experiment~\cite{zou_arXiv_2022}, demonstrating that a fully gapped superconductor becomes a nodal phase by substituting S into single-layer FeSe/SrTiO$_3$.

\section{Model Hamiltonian}
\label{sec-two}
	
In this section, we first discuss the normal-state Hamiltonian that will be used throughout this work for an electronic system consisting of both spin and two locally degenerate atomic orbitals (e.g.,~$d_{xz}$ and $d_{yz}$) on a 2D square lattice. We assume each unit cell contains only one atom, so there is no sublattice degree of freedom. The orbital degeneracy can be reduced by applying the in-plane lattice strain because the original $C_{4v}$ point group is reduced down to its subgroup $C_{2v}$ for strain $\sigma_{10}, \sigma_{01}$ or $\sigma_{11}$ (A more generic strain would reduce the symmetry directly to $C_2$). Here $\sigma_{n_1n_2}$ represents the strain tensor whose form will be given later. We will apply the symmetry analysis to construct the strained Hamiltonian in the spirit of ${\bf k}\cdot{\bf p}$ theory. Then, we discuss the pairing Hamiltonian and the corresponding classification of spin-singlet pairing symmetries including non-unitary pairing states. The strain effect is also investigated on the superconducting pairing symmetries based on a weak-coupling scheme~\cite{zeng_arxiv_2021b}.

\begin{figure*}[t]
\centering
\includegraphics[width=0.8\linewidth]{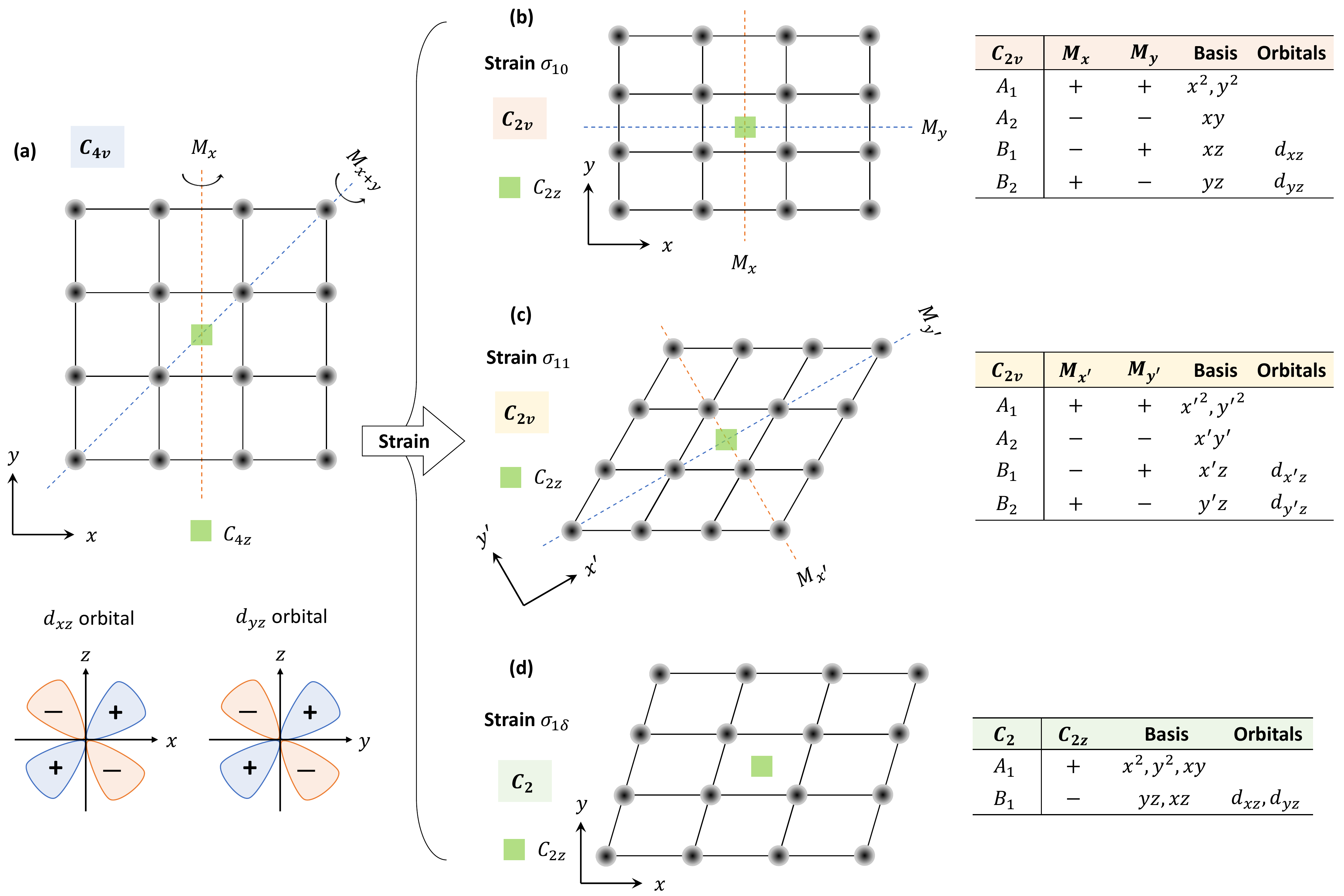}
\caption{
The strain effect on a two-dimensional square lattice. In the absence of lattice strain, (a) shows the square lattice owing the $C_{4v}$ point group that is generated by $C_{4z}$ and $M_x$. We consider the normal-state Hamiltonian with $d_{xz},d_{yz}$-orbitals. Inversion symmetry (${\mathcal I}$) is broken by growing crystal samples on an insulating substrate.
The in-plane strain effects on the square lattice are illustrated in (b-d) for applied strain along different directions.
(b) shows that the $\hat{x}$ or $\hat{y}$-axis strain breaks the square lattice into the rectangular lattice with two independent mirror reflection symmetries $M_x$ and $M_y$, obeying the subgroup $C_{2v}$ of $C_{4v}$. The $C_{2v}$ point group contains four one-dimensional irreducible representations (irrep.) $A_1, A_2, B_1, B_2$. 
(c) shows that the strain along the $\hat{x}+\hat{y}$-direction also reduces the $C_{4v}$ down to $C_{2v}$.
(d) represents a general case, where the subgroup $C_2$ is preserved that only has $A_1$ and $B_1$ irreps.
}
\label{fig:lattice}
\end{figure*}

\subsection{Normal-state Hamiltonian}

In this subsection, we construct the two-orbital normal-state Hamiltonian ${\cal H}_0({\bf k})$ with lattice strain-induced symmetry-breaking terms. Before that, We first show ${\cal H}_0({\bf k})$ in the absence of external lattice strains. For a square lattice as illustrated in Fig.~\ref{fig:lattice} (a), it owns the $C_{4v}$ point group that is generated by two symmetry operators: a fourfold rotation symmetry around the $\hat{z}$-axis $C_{4z}: (x,y)\to(y,-x)$ and a mirror reflection about the $\hat{y}-\hat{z}$ plane $M_x: (x,y)\to(-x,y)$. Other symmetries can be generated by multiplications, such as the mirror reflection about the $(\hat{x}+\hat{y})-\hat{z}$ plane $M_{x+y}: (x,y)\to(y,x)$ is given by $C_{4z}\times M_x$. In the absence of Rashba spin-orbit coupling (SOC), the system also harbors inversion symmetry ${\cal I}$, enlarging the symmetry group to $D_{4h}=C_{4v}\otimes\{E, {\cal I}\}$. In the spirit of ${\bf k}\cdot{\bf p}$ expansion around the $\Gamma$ point or the $M$ point,
we consider a two-orbital system described by the inversion-symmetric Hamiltonian in two dimensions (2D), 
\begin{equation}	
\label{Eq::normal_H}
\mathcal{H}_0(\mathbf{k}) = \epsilon(\mathbf{k})\tau_0\sigma_0 + \lambda_{soc}\tau_2\sigma_3  +\lambda_o[ \mathbf{g}_o(\mathbf{k})\cdot \bm{\tau}]\sigma_0,
\end{equation}
where the basis is made of $\{d_{xz},d_{yz}\}$-orbitals $\psi_{\mathbf{k}}^\dagger=(c_{d_{xz},\uparrow}^\dagger(\mathbf{k}),  c_{d_{xz},\downarrow}^\dagger(\mathbf{k}),c_{d_{yz},\uparrow}^\dagger(\mathbf{k}),c_{d_{yz},\downarrow}^\dagger(\mathbf{k}))$. Here $c^\dagger$ is the creation operator of electrons, $\bm{\tau}$ and $\bm{\sigma}$ are Pauli matrices acting on the orbital and spin subspace, respectively, and $\tau_0$, $\sigma_0$ are 2-by-2 identity matrices. Besides, $\epsilon(\mathbf{k}) = -(k_x^2+k_y^2)/2m -\mu$ is the band energy measured relative to the chemical potential $\mu$, $m$ is the effective mass, $\lambda_{soc}$ is the atomic SOC~\cite{lee_prb_2010,Clepkens2,boker2019quasiparticle} and $\lambda_o$ characterizes the strength of orbital hybridization. 
This model could describe the two hole pockets of iron-based superconductors~\cite{raghu_prb_2008,yi_review_2017}. Moreover, the first two components of $\mathbf{g}_o(\mathbf{k})$ are for the inter-orbital hopping term, while the third term is for the anisotropic effective mass, explained below in detail.

The $C_{4v}$ (or $D_{4h}$) point group restricts $\mathbf{g}_o(\mathbf{k}) = ( a_o k_xk_y, 0 , k_x^2-k_y^2)$, where $a_o=2$ is a symmetric case that increases the $C_{4z}$ to a continues rotational symmetry about the $\hat{z}$-axis. To be precise, the $g_1$-term, $2\lambda_o k_x k_y \tau_1 \sigma_0$, is attributed to the inter-orbital hopping integral along the $\pm\hat{x}\pm\hat{y}$ directions, 
\begin{align}
\begin{split}
\frac{\lambda_o}{2} (  &c_{d_{xz},\sigma}^\dagger(i_x,i_y) \; c_{d_{yz},\sigma}(i_x+1,i_y+1) \\
+& c_{d_{xz},\sigma}^\dagger(i_x,i_y) \; c_{d_{yz},\sigma}(i_x-1,i_y-1) \\
-& c_{d_{xz},\sigma}^\dagger(i_x,i_y) \; c_{d_{yz},\sigma}(i_x+1,i_y-1) \\
-& c_{d_{xz},\sigma}^\dagger(i_x,i_y) \; c_{d_{yz},\sigma}(i_x-1,i_y+1) 
+ \text{h.c.} ),
\end{split}
\end{align}
where $(i_x,i_y)$ represents the lattice site. In addition, the $g_3$-term, $\lambda_o(k_x^2-k_y^2)\tau_3\sigma_0$, causes the anisotropic effective masses. For example, the effective mass of the $d_{xz}$ orbital is $\frac{1}{1/m-2\lambda_o}$ along the $\hat{x}$-axis while that is $\frac{1}{1/m+2\lambda_o}$ along the $\hat{y}$-axis. This means that the hopping integrals are different along $\hat{x}$ and $\hat{y}$ directions,
\begin{align}
 \begin{split}
  & (\frac{1}{2m} - \lambda_o)
 c_{d_{xz},\sigma}^\dagger(i_x,i_y) \; 
 c_{d_{xz},\sigma}(i_x+1,i_y)  \\
+ & (\frac{1}{2m} + \lambda_o)
 c_{d_{xz},\sigma}^\dagger(i_x,i_y) \; 
 c_{d_{xz},\sigma}(i_x,i_y+1) \\
+ & (\frac{1}{2m} + \lambda_o)
 c_{d_{yz},\sigma}^\dagger(i_x,i_y) \; 
 c_{d_{yz},\sigma}(i_x+1,i_y)  \\
+ & (\frac{1}{2m} - \lambda_o)
 c_{d_{yz},\sigma}^\dagger(i_x,i_y) \; 
 c_{d_{yz},\sigma}(i_x,i_y+1) + \text{h.c.}.
 \end{split}
\end{align}
In this work, we focus on a negative effective mass case by choosing $1/m \pm2\lambda_o>0$. However, using a positive effective mass does not change our main conclusion. 
Moreover, our results can be generally applied to other systems with two orbitals $p_x,p_y$, once it satisfies the $C_{4v}$ point group.

The time-reversal symmetry operator is presented as $\mathcal{T}=i\tau_0\sigma_2\mathcal{K}$ with $\mathcal{K}$ being the complex conjugate. And the inversion symmetry is presented as $\mathcal{I}=\tau_0\sigma_0$. It is easy to show Eq.~\eqref{Eq::normal_H} is invariant under both $\mathcal{T}$ and $\mathcal{I}$. However, inversion can be broken by growing the sample on insulating substrates, the asymmetric Rashba SOC is described by
\begin{align}
	\mathcal{H}_R({\bf k}) = \lambda_R\tau_0 [ \mathbf{g}_R(\mathbf{k})\cdot\bm{\sigma}],
\end{align}
where $\lambda_R$ is the strength of the Rashba SOC with $\mathbf{g}_R(\mathbf{k})=(-k_y,k_x,0)$ as required by the $C_{4v}$ point group.

Next, we consider the lattice strain effect on the two-dimensional crystal with a square lattice, as summarized in Fig.~\ref{fig:lattice} (b-d). The in-plane strain effect is characterized by the 2-by-2 strain tensor $\sigma$ whose elements are defined as $\sigma_{ij} = \frac{1}{2} \left( \partial_{x_i}u_j + \partial_{x_j}u_i \right)$, where $u_i$ is the displacement at $\mathbf{r}$ along the $\hat{e}_i$ direction. Even though it is an abuse of notation, it should be self-evident that the $\sigma$ here does not represent the Pauli matrices. The strain tensor $\sigma$ can be parametrized as the following
\begin{align}
\sigma_{\phi} = \begin{pmatrix}
\cos^2\phi & \cos\phi\sin\phi \\
\cos\phi\sin\phi & \sin^2\phi
\end{pmatrix},
\end{align}
where $\phi$ is the polar angle with respect to the $\hat{x}$-axis. For the $\phi=0$ ($\pi/2$) case, the compressive or tensile strain applied along the $\hat{x}$-axis ($\hat{y}$-axis) makes the square lattice as a rectangular lattice, as illustrated in Fig.~\ref{fig:lattice} (b). And the $\phi=\pi/4$ case is for the shear strain along the $(\hat{x}+\hat{y})$-direction in Fig.~\ref{fig:lattice} (c). All the above cases reduce the $C_{4v}$ point group into its subgroup $C_{2v}$ that is generated by two independent mirror reflections.  Otherwise, it is generally reduced to $C_2$. The irreducible representations for $C_{2v}$ and $C_2$ are shown in Fig.~\ref{fig:lattice} (b-d). Based on the standard symmetry analysis, to the leading order, the strained Hamiltonian is given by 
\begin{align}
	\mathcal{H}_{str} = t_{str} [\sin(2\phi) \tau_1 + \cos(2\phi) \tau_3] \sigma_0,
 \label{eq-strain}
\end{align}
where both $t_{str}$ and $\phi$ can be controlled in experiments~\cite{guo_np_2022}. And $\mathcal{H}_{str}$ can be absorbed into the ${\bf g}_o$-vector in Eq.~\eqref{Eq::normal_H}, renormalizing the orbital hybridization as expected. Furthermore, one can check that $\mathcal{H}_{str}$ preserves both $\mathcal{T}$ and $\mathcal{I}$, but explicitly breaks the $C_{4z}=i\tau_2 e^{i\frac{\pi}{4}\sigma_3}$ because of $[{\cal H}_{str}, C_{4z}]\neq0$. Interestingly, the orbital texture on the Fermi surface can be engineered by strain, and its effect on superconducting pairing symmetries is briefly discussed in the Appendix~\ref{app-strain-effect}.

Therefore, a strained normal-state Hamiltonian is
\begin{align}
\label{eq-totl-normal-ham}
\mathcal{H}_N(\mathbf{k}) = \mathcal{H}_0(\mathbf{k}) + \mathcal{H}_R(\mathbf{k}) + \mathcal{H}_{str},
\end{align}
which will be used throughout this work. The Rashba SOC induced spin-splitting bands are considered only when we discuss the topological superconducting phases in section~\ref{sec-four} and \ref{sec-five}, even though the normal-state Hamiltonian $\mathcal{H}_N(\mathbf{k})$ is topologically trivial.
For the superconducting states, we focus on the inversion symmetric pairings (i.e.,~spin-singlet $s$-wave pairing) and their response to applied strains or in-plane magnetic fields.

\begin{figure}[t]
\centering
\includegraphics[width=\linewidth]{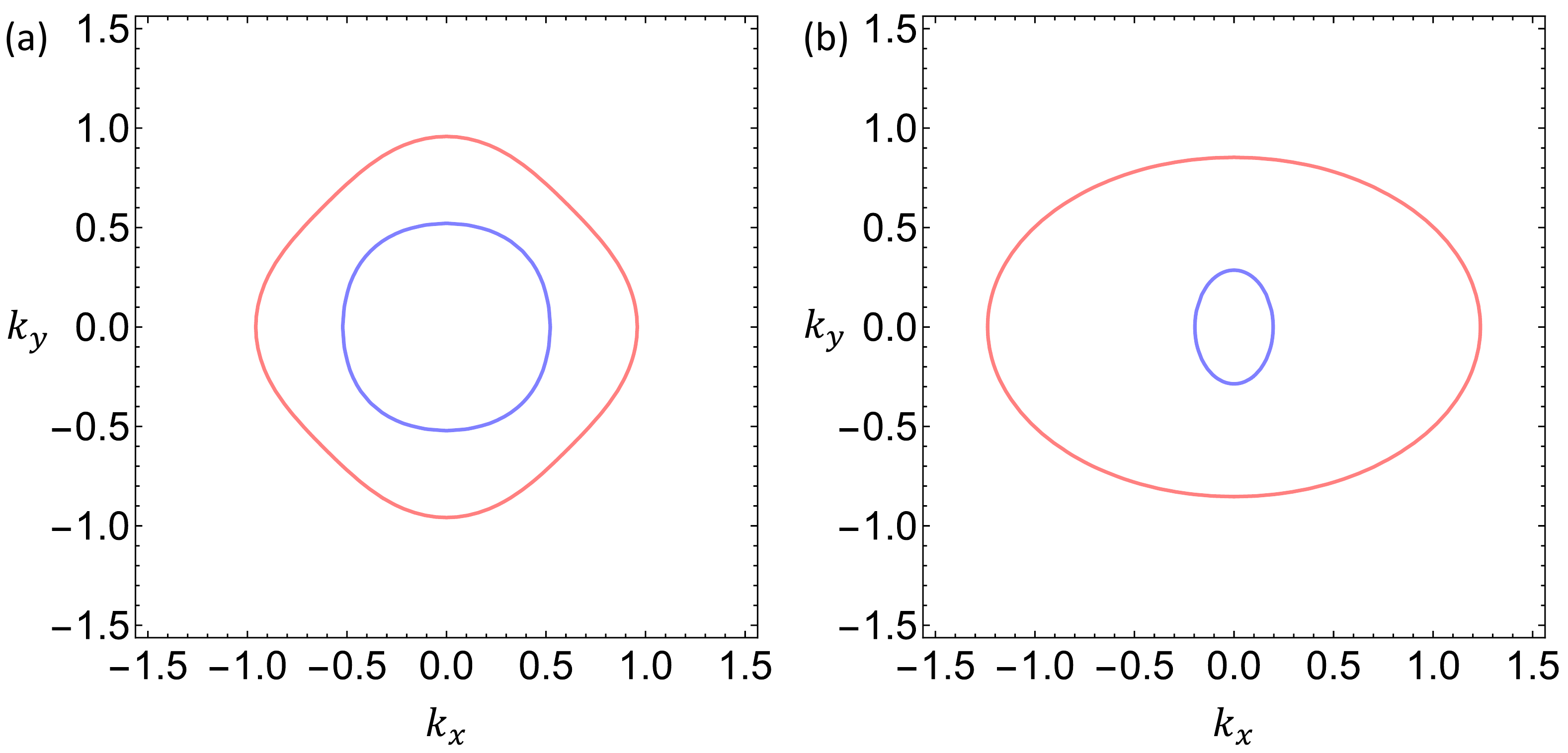}
\caption{
The lattice strain effect on the Fermi surfaces of the normal-state Hamiltonian without Rashba SOC. (a) shows the two Fermi surfaces without lattice strain (i.e.,~$t_{str}=0$), thus $C_{4z}$-symmetric energy contours are formed. (b) shows the breaking of $C_{4z}$ by lattice strain with $t_{str}=0.4$ and $\phi=0$, only $C_{2z}$-symmetric energy contours appear. Other parameters used here are $m=0.5, a_0=1, \lambda_o=0.4$, $\lambda_R=0$ and $\mu=-0.5$.
}
\label{fig:FermiSurf}
\end{figure}

In the absence of Rashba SOC, the band structures of ${\cal H}_N({\bf k})$ in Eq.~\eqref{eq-totl-normal-ham} are given by
\begin{align}
 E_\pm({\bf k}) = -\frac{1}{2m}(k_x^2+k_y^2) \pm \sqrt{\lambda_{soc}^2+\tilde{g}_1^2 + \tilde{g}_3^2},
 \label{eq-energy-band}
\end{align}
where we define the strained orbital hybridization $\tilde{\bf{g}}$ vector with $\tilde{g}_1 = a_o\lambda_o k_xk_y + t_{str}\sin(2\phi)$ and $\tilde{g}_3=\lambda_o(k_x^2 - k_y^2) + t_{str}\cos(2\phi)$. Each band has two-fold degeneracy, enforced by the presence of both ${\cal T}$ and ${\cal I}$. At the $\Gamma$ point ($k_x=k_y=0$), $E_\pm ^ \Gamma = \pm \sqrt{\lambda_{soc}^2 + t_{str}^2}$. The two Fermi surfaces with and without strain are numerically calculated and shown in Fig.~\ref{fig:FermiSurf}, where we choose $\mu < -\sqrt{\lambda_{soc}^2 + t_{str}^2}$. These are two hole pockets because of the negative effective mass of both orbitals. The Fermi surfaces in Fig.~\ref{fig:FermiSurf} (a) are $C_4$-symmetric ($t_{str}=0$), while those in Fig.~\ref{fig:FermiSurf} (b) are only $C_2$-symmetric due to the symmetry breaking of lattice strains. Please note that there is only one Fermi surface when $\vert\mu\vert<\sqrt{\lambda_{soc}^2 + t_{str}^2}$, which is a necessary condition to realize topological superconductors as we will discuss in Sec.~\ref{sec-four}.

\subsection{Review of singlet-triplet mixed pairings}

Before discussing the possible unconventional pairing symmetry for ${\cal H}_N$ in Eq.~\eqref{eq-totl-normal-ham}, we briefly review both unitary and non-unitary gap functions for a single-band SC in the absence of inversion symmetry. In this case, a singlet-triplet mixed pairing potential is given by
\begin{align}
\Delta(\mathbf{k}) = \left\lbrack \Delta_s\psi_s(\mathbf{k})\sigma_0 +  \Delta_t (\mathbf{d}_s(\mathbf{k})\cdot \bm{\sigma}) \right\rbrack (i\sigma_2), 
\end{align}
where $\bm{\sigma}$ are Pauli matrices acting in the spin subspace. Here $\psi_s(\mathbf{k})$ represents even-parity spin-singlet pairings and the odd-parity $\mathbf{d}_s(\mathbf{k})$ is for the spin-triplet $d_s$-vector.
Physically, the unitary SC has only one superconducting gap like in the conventional BCS theory, while a two-gap feature comes into being by the non-unitary pairing potential. More explicitly, the unitary or non-unitary is defined by whether the following is proportional to the identity matrix $\sigma_0$:
\begin{equation}\label{eq-non-unitary-spin-triplet}
\begin{split}
&\Delta(\mathbf{k}) \Delta^\dagger(\mathbf{k}) =\vert\Delta_s\vert^2\psi_s^2+\vert\Delta_t\vert^2\vert\mathbf{d}_s\vert^2\\
&+2\mathrm{Re}\lbrack \Delta_s\Delta_t^*\psi_s\mathbf{d}_s^* \rbrack \cdot\bm{\sigma} 
+i\vert\Delta_t\vert^2(\mathbf{d}_s\times\mathbf{d}_s^*)\cdot\bm{\sigma},
\end{split}
\end{equation}
Therefore, Eq.~\eqref{eq-non-unitary-spin-triplet} gives rise to a possible classification by assuming a non-vanishing $\Delta_s \in \mathds{R}$ and a proper choice of a global phase.
In principle, there are four possible phases, including the TRI non-unitary SCs ($\Delta_t\in \mathds{R} , \mathbf{d}_s \in \mathds{R}$), the TRB unitary SCs ($\Delta_t \sim i , \mathbf{d}_s \in \mathds{R}$), and the TRB non-unitary SCs ($\Delta_t \in \mathds{R} , \mathbf{d}_s \in \mathds{C} $). On the other hand, the TRI unitary SCs are achieved only with $\Delta_s=0$ or $\Delta_t=0$ and real $\mathbf{d}_s$, meaning a purely spin-singlet SC or a purely spin-triplet SC. These states might be distinguished in experiments, for example, the TRB unitary pairing state might induce a spontaneous magnetization with the help of Rashba spin-orbit coupling~\cite{hu_prb_2021}, which can be detected by $\mu$SR~\cite{shang_sa_2022}.

As we know, the spin-singlet pairings do not coexist with the spin-triplet pairings in the presence of inversion symmetry (e.g.~centrosymmetric SCs). Roughly speaking, it seems out of the question to realize non-unitary pairing states in purely spin-singlet SCs. However, this is a challenge but not an impossibility for an SC with multi-orbitals, which is one of the aims of this work. In the following, we will discuss how to generalize the classification of TRI or TRB and unitary or non-unitary pairing states to a spin-singlet SC with two atomic orbitals in the presence of inversion symmetry. The four cases are summarized in Table~\ref{table1}.

 \begin{table}[t]
\centering
\includegraphics[width=\linewidth]{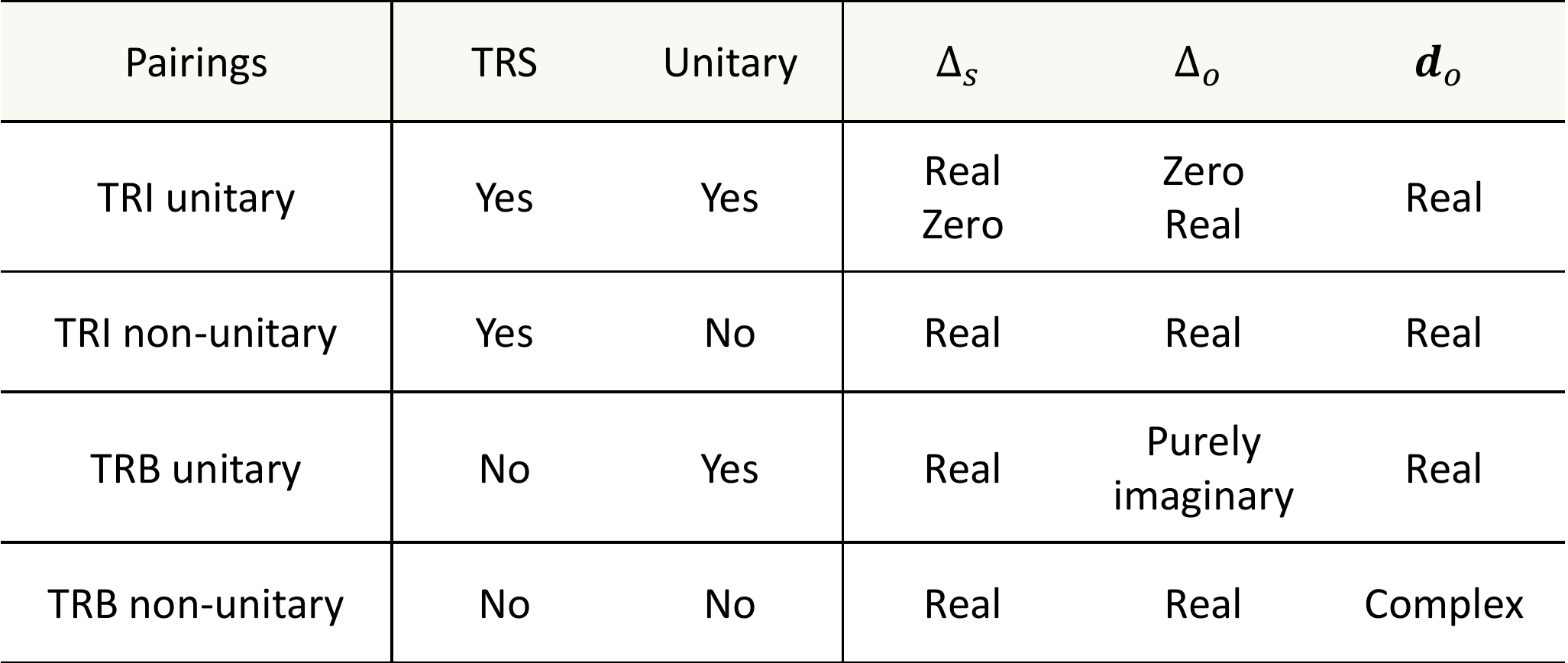}
\caption{
The four pairing states classified by time-reversal symmetry and unitary for a spin-singlet superconductor with both orbital-independent pairing $\Delta_s$ and orbital-dependent pairing $\Delta_o$ and ${\bf d}_o$.
}
\label{table1}
\end{table}

\subsection{Pairing Hamiltonian of a two-orbital model}
	
We now consider the pairing Hamiltonian for ${\cal H}_N$ in Eq.~\eqref{eq-totl-normal-ham}. By ignoring the fluctuations, the mean-field pairing Hamiltonian is generally given by,
\begin{align}
\mathcal{H}_{\Delta} = \sum_{\mathbf{k},s_1a,s_2b}
\Delta_{s_1,s_2}^{a,b}(\mathbf{k}) 
c^\dagger_{as_1}(\mathbf{k})  c^\dagger_{bs_2}(-\mathbf{k}) +\text{h.c.},
\end{align}
where $s_1, s_2$ are index for spin and $a,b$ are for orbitals. 
As studied in Ref.~\cite{zeng_arxiv_2021b}, the orbital-triplet pairing is robust even against orbital hybridization and electron-electron interactions. Thus, we consider both orbital-independent and orbital-dependent pairings for generality.
In analogy to spin-triplet SCs, we use an orbital $\mathbf{d}_o(\mathbf{k})$-vector for the spin-singlet orbital-dependent pairing potential~\cite{ong_prl_2013}, which takes the generic form 
\begin{align}\label{eq-J-2-pairing}
\Delta_{tot}(\mathbf{k})=\left\lbrack \Delta_s\Psi_s(\mathbf{k})\tau_0 + \Delta_o ( \mathbf{d}_o(\mathbf{k})\cdot\bm{\tau} ) \right\rbrack (i\sigma_2),
\end{align} 
where $\Delta_s$ and $\Delta_o$ are pairing strengths in orbital-independent and orbital-dependent channels, respectively. The Fermi statistics requires that $\Psi_s(\mathbf{k}) = \Psi_s(-\mathbf{k})$, while the three components of $\mathbf{d}_o$ satisfy $d_o^{1,3}(\mathbf{k})=d_o^{1,3}(-\mathbf{k})$ and $d_o^2(\mathbf{k})=-d_o^2(-\mathbf{k})$. 
Namely, $d_o^2(\mathbf{k})$ represents odd-parity spin-singlet orbital-singlet pairings and the others are for even-parity spin-singlet orbital-triplet pairings.
The pairing potential presented in this form is quite convenient, similar to the spin-triplet case~\cite{frigeri_prl_2004,ong_prl_2013,ong_pnas_2016,zeng_arxiv_2021b}. The benefits of this form in Eq.~\eqref{eq-J-2-pairing} will be shown when we discuss the mixture of orbital-independent and orbital-dependent pairings.
Combining Eq.~\eqref{eq-J-2-pairing} with Eq.~\eqref{Eq::normal_H}, the Bogoliubov-de-Gennes (BdG) Hamiltonian is
\begin{equation}\label{eq-bdg-ham}
\mathcal{H}_{\text{BdG}}({\bf k})=\begin{pmatrix}
\mathcal{H}_N(\mathbf{k}) & \Delta_{tot}(\mathbf{k})\\
\Delta_{tot}^\dagger(\mathbf{k})& -\mathcal{H}_N^\ast(-\mathbf{k})
\end{pmatrix},
\end{equation}
which is based on the Nambu basis $(\psi_{\mathbf{k}}^\dagger,\psi_{-\mathbf{k}}^T)$. Same with the spin case in Eq.~\eqref{eq-non-unitary-spin-triplet}, the non-unitarity of a spin-singlet pairing potential defined in Eq.~\eqref{eq-J-2-pairing} is determined by whether $\Delta_{tot}(\mathbf{k}) \Delta_{tot}^\dagger(\mathbf{k})$ is proportional to an identity matrix. More explicitly we have
\begin{equation}
\begin{split}
& \Delta_{tot}(\mathbf{k}) \Delta_{tot}^\dagger(\mathbf{k}) =\vert\Delta_s\vert^2\psi_s^2\tau_0
\sigma_0+\vert\Delta_o\vert^2\vert\mathbf{d}_o\vert^2\tau_0\sigma_0\\
&+2\mathrm{Re}\vert \Delta_s\Delta_o^*\psi_s\mathbf{d}_o^*\vert \cdot\bm{\tau}\sigma_0
+i\vert\Delta_o\vert^2(\mathbf{d}_o\times\mathbf{d}_o^*)\cdot\bm{\tau}\sigma_0,
\end{split}
\end{equation}
which could also exhibit four general possibilities: time-reversal-invariant (TRI) or time-reversal-breaking (TRB) and unitary or non-unitary SCs, with a simple replacement $\{\Delta_t,\mathbf{d}_s\} \to \{\Delta_o, \mathbf{d}_o \}$. 
In the absence of band splittings, i.e., $\lambda_{soc}=\lambda_o=\lambda_R=t_{str}=0$ as an illustration, the superconducting excitation gaps on the Fermi surfaces of a TRI unitary SC are
\begin{align}
E_{\kappa,\nu} (\mathbf{k}) = \kappa\sqrt{ \epsilon^2(\mathbf{k})  + (\Delta_s \psi_s(\mathbf{k})  +\nu \Delta_o\vert\mathbf{d}_o\vert)^2  },
\end{align}
with $\kappa,\nu=\pm$. It is similar to the superconducting gaps for non-unitary spin-triplet SCs \cite{sigrist_rmp_1991}. Moreover, the two-gap feature indicates the non-unitarity of the superconducting states, which implies the possibility of a nodal SC as long as $\Delta_s \psi_s(\mathbf{k}) \pm \Delta_o\vert\mathbf{d}_o\vert=0$ is satisfied on the Fermi surfaces. And, the nodal quasi-particle states can be experimentally detected by measuring specific heat, London penetration depths, $\mu$SR, NMR, etc. As a result, this provides possible evidence to get a sight of TRI non-unitary phases in real materials (e.g.~centrosymmetric SCs). Furthermore, the above conclusion is still valid when we turn on $\lambda_{soc}$, $\lambda_o$, and $t_{str}$.

\section{The Pauli limit violation: a large in-plane upper critical field}
\label{sec-three}

In this section, we study the Pauli limit violation for the spin-singlet TRI non-unitary SC against an in-plane magnetic field (e.g.~$H_{c2,\parallel}>H_P$). For a 2D crystalline SC or a thin film SC, the realization of superconducting states that are resilient to a strong external magnetic field has remained a significant pursuit, namely, the pairing mechanism can remarkably enlarge the in-plane upper critical field. Along this crucial research direction, one recent breakthrough has been the identification of ``Ising pairing'' formed with the help of Ising-type spin-orbit coupling (SOC), which breaks the SU(2) spin rotation and pins the electron spins to the out-of-plane direction. Depending on whether the inversion symmetry is broken or not by the Ising-type SOC, the Ising pairing is classified as Type \uppercase\expandafter{\romannumeral1} (broken) and Type \uppercase\expandafter{\romannumeral2} (preserved) Ising superconductivity, where the breaking of Cooper pairs is difficult under an in-plane magnetic field.

To demonstrate the underlying physics, in the following, we consider the interplay between atomic SOC $\lambda_{soc}\neq0$ and spin-singlet TRI non-unitary pairing state. Thus, we consider the pairing potential
\begin{align}
\label{eq-delta-TRI-nonunitary-main}
\Delta_{tot}=\left\lbrack \Delta_s\tau_0+\Delta_o(d_o^1\tau_1+d_o^3\tau_3)\right\rbrack (i\sigma_2),
\end{align}
where $\Delta_s$, $\Delta_o$, $d_o^1$, and $d_o^3$ are all real constant. This can be realized once we have on-site attractive interactions in both orbital channels. Another reason for studying the atomic SOC is that it is not negligible in many real materials. 
It is interesting to note that the strength of SOC can be tuned in experiments, for example, by substituting S into single-layer FeSe/SrTiO$_3$~\cite{zou_arXiv_2022} or growing a superconductor/topological insulator heterostructure~\cite{yi_nm_2022}.

Without loss of generality, the direction of the magnetic field can be taken to be the $x$-direction, i.e., $\mathbf{H}=(H_x,0,0)$ with $H_x\ge0$. Therefore, the normal Hamiltonian becomes
\begin{equation}\label{eq-ham-hx}
\mathcal{H}_N(\mathbf{k}) + h\tau_0\sigma_1,
\end{equation}
where the first part is given by Eq.~\eqref{eq-totl-normal-ham} and $h=\tfrac{1}{2}g\mu_BH_x$ is the Zeeman energy with $g=2$ the electron's g-factor. To explicitly investigate the violation of the Pauli limit for the spin-orbit coupled SCs, we calculate the in-plane upper critical magnetic field normalized to the Pauli-limit paramagnetic field $H_{c2,\parallel}/H_P$ as a function of the normalized temperature $T_c/T_0$, by solving the linearized gap equation. Here $H_P=1.86T_0$ represents the Pauli limit with $T_0$ the critical temperature in the absence of an external magnetic field.

Following the standard BCS decoupling scheme~\cite{zeng_arxiv_2021b}, we first solve $T_c$ for the orbital-independent pairing channel by solving the linearized gap equation, $v_0\chi_s(T)-1 = 0$, where $v_{0}$ is effective attractive interaction and the superconductivity susceptibility $\chi_s(T)$ is defined by
\begin{align}\label{eq-sus-chiS}
\chi_s(T) = -\frac{1}{\beta}\sum_{\mathbf{k},\omega_n} \text{Tr}\Big{\lbrack}  G_e(\mathbf{k},i\omega_n)  G_h(-\mathbf{k},i\omega_n) \Big{\rbrack},
\end{align}
where the conventional s-wave pairing with $\psi_s(\mathbf{k})=1$ is considered for Eq.~\eqref{eq-delta-TRI-nonunitary-main}. Here $G_e(\mathbf{k},i\omega_n)=[i\omega_n-\mathcal{H}_0(\mathbf{k})]^{-1}$ is the Matsubara Green's function for electrons and that for holes is defined as $G_h(\mathbf{k},i\omega_n)=-\sigma_2G_e^\ast(\mathbf{k},i\omega_n)\sigma_2$. Here $\beta=1/k_BT$ and $\omega_n=(2n+1)\pi/\beta$ with $n$ integer. 
Likewise, for the orbital-dependent pairing channels, the superconductivity susceptibility $\chi_o(T)$ is defined as
\begin{align}\label{eq-sus-chiT}
\begin{split}
\chi_o(T) = -\frac{1}{\beta}\sum_{\mathbf{k},\omega_n} \text{Tr}\Big{\lbrack} &(\mathbf{d}_o(\mathbf{k})\cdot\bm{\tau})^\dagger G_e(\mathbf{k},i\omega_n)  \\ 
\times &(\mathbf{d}_o(\mathbf{k})\cdot\bm{\tau}) G_h(-\mathbf{k},i\omega_n) \Big{\rbrack}, 
\end{split}
\end{align}
where the orbital-dependent pairing ($A_g$ representation) with the vector-form as ${\bf d}_o=(d_o^1,0,d_o^3)$ for Eq.~\eqref{eq-delta-TRI-nonunitary-main} is used for the $T_c$ calculations. 
However, the momentum-dependent ${\bf d}_o$-vector does not affect the formalism and main results, as we will discuss in the appendix~\ref{app-strain-effect}.
The coupling between orbital-independent and orbital-dependent channels leads to a high-order correction ($\sim\lambda^2k_F^2/\mu^2$, with $\lambda$ being the coupling strength of the effective $\tilde{\bf g}$ in the Hamiltonian representing orbital hybridization and strain ), which can be ignored once $\lambda\ll\mu/k_F$.

\begin{figure*}[t]
\centering
\includegraphics[width=\textwidth]{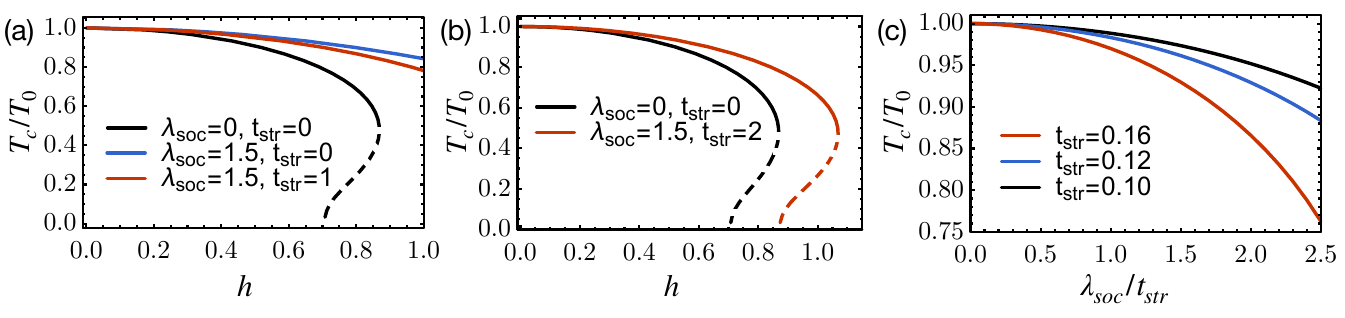}
\caption{The pair-breaking effects. (a) A significant Pauli limit violation is due to the atomic SOC for the orbital-independent pairing with $\Delta_s=1$. 
However, the lattice strain might slightly suppress the $H_{c2}$ by comparing the blue and red curves. (b) A weak Pauli limit violation due to the atomic SOC for the orbital-dependent pairing with $\Delta_o=1$. (c) The suppression of $T_c$ by atomic SOC for orbital-dependent pairing at zero external magnetic fields with $\Delta_o=1$. 
For the three figures here, we have set the strain parameter $\phi=\frac{\pi}{8}$, i.e. $\tilde{\mathbf{g}}=(\frac{\sqrt{2}}{2},0,\frac{\sqrt{2}}{2})$. 
}
\label{fig-sec3}
\end{figure*}

\subsection{Type \uppercase\expandafter{\romannumeral2} Ising superconductivity}

In this subsection, we first consider the orbital-independent pairing state (i.e.,~$\Delta_s\neq0$ and $\Delta_o=0$) and show it is a Type II Ising SC protected from the out-of-plane spin polarization by the atomic SOC $\lambda_{soc}\tau_2\sigma_3$.
To demonstrate that, one generally needs to investigate the effects of atomic SOC on the SC $T_c$ as a function of the in-plane magnetic field $h$ based on Eq.~\eqref{eq-sus-chiS}, in the presence of both orbital hybridization $\lambda_o$ and strain $t_{str}$. As defined in Eq.~\eqref{eq-energy-band}, the effects of orbital hybridization and lattice strain on the system can be captured by an effective $\Tilde{\bf g}\equiv  (a_o\lambda_o k_xk_y + t_{str}\sin(2\phi), \lambda_o(k_x^2 - k_y^2) + t_{str}\cos(2\phi))$. The case with $t_{str}=0$ has been studied in Ref.~\cite{wang_prl_2019}, however, the strain effect on the type \uppercase\expandafter{\romannumeral2} Ising SC has not been explored yet. To reveal the pure role of lattice strains, we consider $k_F$ to be close to the $\Gamma$-point so that the $k$-dependent hybridization part is dominated by the strain part for generic $\phi$. Therefore, we focus on $\Tilde{\bf g}=t_{str}(\sin2\phi,\cos2\phi)$ in the following discussions.

After a straightforward calculation (see details in Appendix~\ref{append-Tc-derivation}), the superconductivity susceptibility $\chi_s(T)$ in Eq.~\eqref{eq-sus-chiS} is calculated as
\begin{align}
\chi_s(T) = \chi_0(T)+N_0 f_s(T,\lambda_{soc},t_{str},h),
\end{align}
with $N_0$ is the DOS near the Fermi surface and the pair-breaking term is given by
\begin{align}\label{eq-depair-fs}
\begin{split}
f_s(T,\lambda_{soc},t_{str},h) = \frac{1}{2}\left[\mathcal{C}_0\left(T,\rho_-\right)+\mathcal{C}_0\left(T,\rho_+\right)\right] \\
+\left[\mathcal{C}_0\left(T,\rho_-\right)-\mathcal{C}_0\left(T,\rho_+\right)\right]\left(\frac{\lambda_{soc}^2+t_{str}^2-h^2}{2E_+E_-}\right) ,
\end{split}
\end{align}
where $E_{\pm}\equiv \sqrt{\lambda_{soc}^2+(t_{str}\pm h)^2}$, $\rho_\pm = \frac{1}{2}(E_+ \pm E_-)$, and $\chi_0(T)=N_0\ln\left( \tfrac{2e^\gamma \omega_D}{\pi k_B T} \right)$ is the superconducting susceptibility when $\lambda_{soc},t_{str},h=0$. Here $\gamma=0.57721\cdots$ is the Euler-Mascheroni constant. Furthermore, the kernel function of the pair-breaking term $f_s$ is given by
\begin{align}\label{eq-kernel-c0}
\mathcal{C}_0(T,E)=\text{Re}\left\lbrack \psi^{(0)}(\frac{1}{2}) - \psi^{(0)}(\frac{1}{2}+i\frac{E}{2\pi k_BT}) \right\rbrack,
\end{align}
with $\psi^{(0)}(z)$ being the digamma function. Note that $\mathcal{C}_0(T,E) \le 0$ and it monotonically decreases as $E$ increases, indicating the reduction of $T_c$. Namely, $\mathcal{C}_0(T,E)$ gets smaller for a larger $E$.

We first discuss the simplest case with $\lambda_{soc}=t_{str}=0$, where the pair-breaking function becomes $f_s(T,0,0,h) = \mathcal{C}_0(T,h)$, which just leads to the Pauli limit $H_{c2,\parallel} \approx H_P = 1.86 T_{c}$, as shown in Fig.~\ref{fig-sec3}(a). Furthermore, we turn on $\lambda_{soc}$ while take the $t_{str}\to 0$ limit, the pair-breaking term in Eq.~\eqref{eq-depair-fs} is reduced to
\begin{align}\label{eq-fs-lam0-case}
f_s(T,\lambda_{soc},0,h) = \mathcal{C}_0\left(T,\sqrt{\lambda_{soc}^2+h^2}\right) \frac{h^2}{\lambda_{soc}^2+h^2},
\end{align}
which reproduces the same results of Type II Ising superconductors in Ref.~\cite{wang_prl_2019}. Under a relatively weak magnetic field ($h\ll\lambda_{soc}$), the factor $h^2/(\lambda_{soc}^2+h^2)\ll 1$ leads to  $f_s(T,\lambda_{soc},0,h)\to0$, which in turn induces a large in-plane $H_{c2,\parallel}/H_P$.

Next, we investigate the effect of lattice strain $t_{str}$ on the in-plane upper critical field $H_{c2,\parallel}$. Interestingly, $t_{str}$ would generally instead reduce $H_{c2,\parallel}$. To see it explicitly, we expand the pair-breaking function $f_s$ in Eq.~\eqref{eq-depair-fs} up to the leading order of $t_{str}^2$,
\begin{align}
\begin{split}
f_s( T,\lambda_{soc},t_{str},h ) &\approx f_s( T,\lambda_{soc},0,h ) \\
&+ F(T,\lambda_{soc},h) t_{str}^2 + {\cal O}(t_{str}^4),
\end{split}
\end{align}
where $F(T,\lambda_{soc},h)$ is given in Appendix~\ref{append-Tc-derivation} and we find it is always negative (i.e.,~$F(T,\lambda_{soc},h)<0$). In addition to the first term $f_s( T,\lambda_{soc},0,h )$ discussed in Eq.~\eqref{eq-fs-lam0-case}, the second term $F(T,\lambda_{soc},h) t_{str}^2$ also serves as a pair-breaking effect on $T_c$ at non-zero field. Therefore, the second $\lambda_o$-term further reduces $T_c$, leading to the reduction of the in-plane upper critical field.

We then numerically confirm the above discussions. We solve the linearized gap equation $v_0\chi_s(T)-1=0$ and arrive at $\log(T_c/T_0) = f_s(T_c,\lambda_{soc},t_{str},h)$, from which $T_c/T_0$ is numerically calculated in Fig.~\ref{fig-sec3} (a). Here $T_0$ is the critical temperature at zero external magnetic fields. The Pauli limit corresponds to $T_0(\lambda_{soc}=0,t_{str}=0,h=0)$. 
The non-monotonic behavior of the curves at small $T_c/T_0$ ($\lesssim 0.5$, i.e. dashed line) from solving the linearized gap equation calls for a comment. In the small temperature range, the transition by tuning the field strength becomes the first order supercooling transition~\cite{maki1964}. Here we mainly focus on the solid line part, which is second order and gives the critical field $H_{c2}$.
We see that in general there is a Pauli limit violation for non-zero $\lambda_{soc}$ and $t_{str}$. Furthermore, by comparing the two cases with $\lambda_{soc}=1.5,t_{str}=0$ and $\lambda_{soc}=1.5,t_{str}=1$, we confirm the above approximated analysis. 
We believe the strain effect on the type \uppercase\expandafter{\romannumeral2} Ising SC will be tested in experiments soon.

\subsection{$H_{c2,\parallel}$ for orbital-dependent pairings}

In this subsection, we further study the influence of the atomic SOC $\lambda_{soc}$ on the paramagnetic pair-breaking effect for orbital-dependent pairings (i.e.,~$\Delta_s=0$ and $\Delta_o\neq0$). We find a weak enhancement of the in-plane upper critical field $H_{c2,\parallel}$ compared with the Pauli limit. Following the criteria of the orbital $\mathbf{d}_o$-vector in Ref.~\cite{zeng_arxiv_2021b} (also discussed in the Appendix~\ref{app-strain-effect}), we take $\mathbf{d}_o$ to be parallel to the vector $\tilde{\mathbf{g}}$ by assuming $\lambda_{soc}\ll t_{str}$, which leads to the maximal condensation energy. This would be justified in the next subsection. After a straightforward calculation (see details in Appendix~\ref{append-Tc-derivation}), the superconductivity susceptibility $\chi_o(T)$ in Eq.~\eqref{eq-sus-chiT} is calculated as,
\begin{align}
\chi_o(T) = \chi_0(T)+N_0 f_o(T,\lambda_{soc},t_{str},h),
\end{align}
where the pair-breaking term is given by
\begin{align}\label{eq-depair-fo}
\begin{split}
f_o(T,\lambda_{soc},t_{str},h) = 	\frac{1}{2}\left[\mathcal{C}_0\left(T,\rho_-\right)+\mathcal{C}_0\left(T,\rho_+\right)\right] \\
+\left[\mathcal{C}_0\left(T,\rho_-\right)-\mathcal{C}_0\left(T,\rho_+\right)\right] \left(\frac{t_{str}^2-\lambda_{soc}^2-h^2}{2E_+E_-}\right),
\end{split}
\end{align}
which differs from $f_s(T,\lambda_{soc},t_{str},h)$ for orbital-independent pairings in Eq.~\eqref{eq-depair-fs}. The only difference between them lies in the factor $( t_{str}^2-\lambda_{soc}^2-h^2)/2E_+E_-$, compared with that of $f_s(T,\lambda_{soc},t_{str},h)$ (i.e. $( t_{str}^2+\lambda_{soc}^2-h^2)/2E_+E_-$), which leads to a completely distinct superconducting state, demonstrated as follows.

To understand Eq.~\eqref{eq-depair-fo}, we first discuss the simplest case with $\lambda_{soc}=t_{str}=0$, where the pair-breaking function becomes $f(T,0,0,h) = \mathcal{C}_0(T,h)$, which just leads to the Pauli limit $H_{c2,\parallel} \approx H_P = 1.86 T_{c}$, as shown in Fig.~\ref{fig-sec3}(b). Likewise, when $\lambda_{soc}=0$ and $t_{str}\neq0$, the pair-breaking function again simplifies to $\mathcal{C}_0(T,h)$. Therefore, the Pauli limit of the in-plane upper critical field is not affected by $t_{str}$ itself. 
Physically, this is because spin and orbital degrees of freedom are completely decoupled in this case, and it has also been shown that a similar orbital effect does not suppress $T_c$ when $\mathbf{d}_o\parallel \tilde{\bf g}$ \cite{zeng_arxiv_2021b}, which is what we assumed here.

On the other hand, if we turn on merely the atomic SOC $\lambda_{soc}\neq0$ while keeping $t_{str}=0$, the pair-breaking function is given by
\begin{align}
f_o(T,\lambda_{soc},0,h) = \mathcal{C}_0(T,\sqrt{\lambda_{soc}^2+h^2}),
\end{align}
which leads to the reduction of the upper critical field, i.e., $H_{c2,\parallel} < H_P$, because of $f(T,\lambda_{soc},0,h)  < f(T,0,0,h) <0$. Remarkably, we find that the atomic SOC also plays a similar role of ``magnetic field'' to suppress the orbital-dependent pairing, as discussed in the next subsection. Thus, it does not belong to the family of Ising SCs, which makes the orbital-dependent pairing significantly different from the orbital-independent pairings. Moreover, their different dependence on the in-plane magnetic field might also be tested in experiments, which is beyond this work and left for future work. This also indicates the difference between orbital-triplet SC and spin-triplet SC in responses to Zeeman fields.

However, it is surprising to notice that there is a weak enhancement of the in-plane upper critical field $H_{c2,\parallel}$ for the case with both $t_{str}\neq0$ and $\lambda_{soc}\neq0$. Solving the gap equation $v_0\chi_o(T)-1=0$, we obtain
\begin{equation}
\mathrm{ln}\left(\frac{T_c}{T_0}\right)=f_o(T,\lambda_{soc},t_{str},h).
\end{equation}
Fig.~\ref{fig-sec3} (b) shows how $T_c/T_0$ changes with the applied in-plane magnetic field, where the Pauli limit curve corresponds to $\lambda_{soc},t_{str}=0$. When both the atomic SOC and strain are included, the critical field $H_{c2}$ exceeds the Pauli limit by a small margin. Therefore, a spin-orbit-coupled SC with spin-singlet non-unitary pairing symmetries does not belong to the reported family of Ising superconductivity.

\subsection{Atomic SOC induced zero-field Pauli limit}

As mentioned above, the atomic SOC breaks the spin degeneracy, which generally suppresses the even parity orbital-dependent pairings, in the case with $\Delta_s=0$ and $\Delta_o\neq0$. Thus, the robustness of such pairings in the presence of atomic SOC is the preliminary issue that we need to address. And we find that the spin-singlet orbital-dependent pairing is also prevalent in solid-state systems when the energy scale of atomic SOC is smaller than that of the orbital hybridization or external strain. In this case, we focus on the zero magnetic field limit. Using the general results from the calculations in the previous section, we have
\begin{equation}\label{eq-tc-suppression-soc}
\begin{split}
\mathrm{ln}\left(\frac{T_c}{T_0}\right) &=f_o(T,\lambda_{soc},t_{str},h=0) \\
&=\mathcal{C}_0\left(T, \sqrt{t_{str}^2+\lambda_{soc}^2} \right) \frac{\lambda_{soc}^2}{t_{str}^2+\lambda_{soc}^2},
\end{split}
\end{equation}
where $\mathcal{C}_0(T,E)$ is defined in Eq.~\eqref{eq-kernel-c0}. In the case of $\lambda_{soc}=0$, it can be been that $T_c(t_{str})=T_0(t_{str}=0)$, i.e. the superconducting $T_c$ is not suppressed by stain or the orbital hybridization when the orbital $\mathbf{d}_o$-vector is parallel to $\tilde{\bf g}$ \cite{zeng_arxiv_2021b}. However, in the presence of non-zero atomic SOC $\lambda_{soc}$, the $T_c$ will be suppressed even when $\mathbf{d}_o \parallel \tilde{\bf g}$ is satisfied. Fig.~\ref{fig-sec3} (c) shows the behavior of $T_c$ as a function of the $\lambda_{soc}/t_{str}$ for two different values of $t_{str}$. We see the suppression of $T_c$ as long as $\lambda_{soc}\neq 0$, and the suppression is more prominent when $t_{str}$ is larger.

To understand the suppression of orbital-dependent pairings by the atomic SOC, we take the $t_{str}=0$ limit. Eq.~\eqref{eq-tc-suppression-soc} leads to 
\begin{align}
\mathrm{ln}\left(\frac{T_c}{T_0}\right) =\mathcal{C}_0\left(T, \lambda_{soc} \right),
\end{align}
which implies that $\lambda_{soc}$ plays the same role of ``magnetic field'' that suppresses the $T_c$ of the orbital-dependent pairing states. And $\lambda_{soc}\sim H_P$ roughly measures the zero-field ``Pauli-limit'' of the orbital-dependent pairing states. 
We dub this new effect as zero-field Pauli limit for orbital-dependent pairings induced by the atomic SOC, which can serve as the preliminary analysis of whether orbital-dependent pairings exist or not in real materials by simply calculating $\lambda_{soc}/T_c$.

Motivated by this observation, we notice that the normal Hamiltonian given in Eq.~\eqref{eq-totl-normal-ham} satisfies $[\mathcal{H}_N({\bf k}),\tau_2]=0$ with both $\lambda_o\to0$ and $t_{str}\to 0$. It stands for the U(1) rotation in the orbital subspace. As a result, we can project the normal Hamiltonian $\mathcal{H}_N({\bf k})$ in Eq.~\eqref{eq-totl-normal-ham} into block-diagonal form corresponding to the $\pm 1$ eigenvalues of $\tau_2$ by using the basis transformation 
\begin{align}
\mathcal{U}=\sigma_0\otimes \frac{1}{\sqrt{2}}\begin{bmatrix}1&-i\\1&i \end{bmatrix}.
\end{align} 
The new basis is given by 
\begin{align}
\Tilde{\Psi}^\dagger(\mathbf{k})=(c_{+,\uparrow}^\dagger,c_{+,\downarrow}^\dagger,c_{-,\downarrow}^\dagger,c_{-,\uparrow}^\dagger),
\end{align}
where $c_{\pm,s}^\dagger\equiv\frac{1}{\sqrt{2}}(c_{d_{xz},s}^\dagger\mp i c_{d_{yz},s}^\dagger)$. 
On this basis, the normal Hamiltonian is given by 
\begin{align}
\mathcal{H}_0=\mathcal{H}_0^{+}\oplus \mathcal{H}_0^{-},
\end{align} 
where $\mathcal{H}_0^{\pm}$ are given by
\begin{align}
\mathcal{H}_0^{\pm}=\epsilon(\mathbf{k}) \mp \lambda_{soc}\sigma_3.
\end{align}
Note that the time-reversal transforms $\mathcal{H}_0^{\pm}(\mathbf{k})$ to $\mathcal{H}_0^{\mp}(-\mathbf{k})$. Explicitly, the atomic SOC is indeed a ``magnetic field'' in each subspace, while it switches signs in the two subspaces to conserve TRS.

\begin{figure*}[t]
\centering
\includegraphics[width=\textwidth]{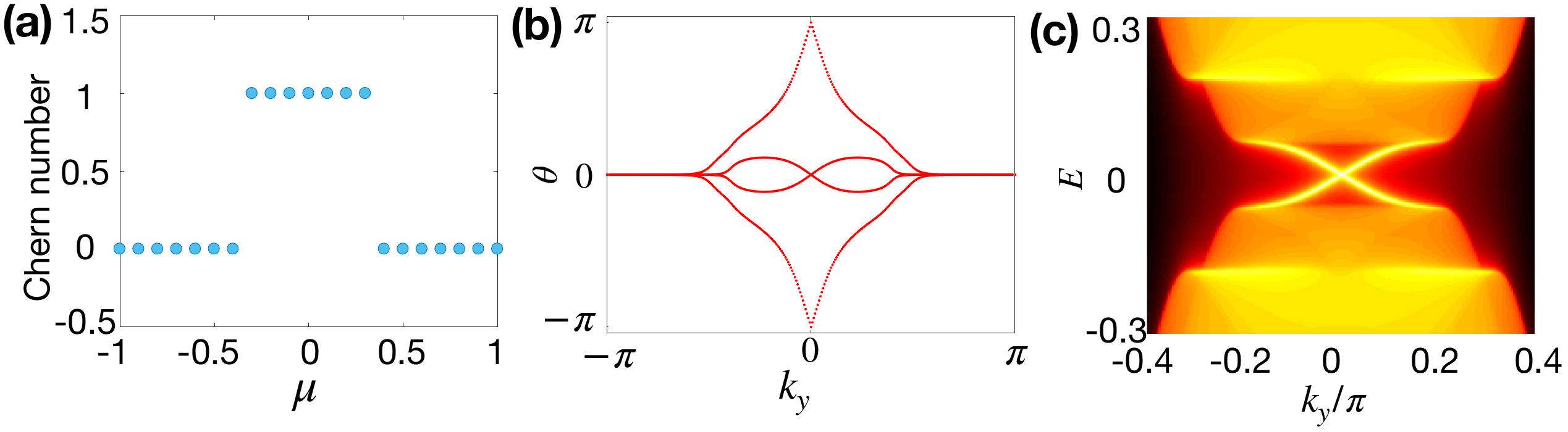}
\caption{Topological helical superconductivity for spin-singlet orbital-dependent pairing in the presence of Rashba SOC. (a) The $Z_2$ index is calculated by decoupling the BdG Hamiltonian into two chiral blocks when $\Delta_s=0$ and $t_{str}=0$. The other parameters used: $m=0.5$, $\mu=0.2$, $\lambda_{soc}=0.4$, $\lambda_R=1$, $\Delta_o=0.1$, $\mathbf{d}_o=(1,0,1)$. (b) The Wilson loop calculation of the $Z_2$ invariant for $\Delta_s=0.05$, $t_{str}=0.1$, $\phi=\frac{\pi}{8}$ and $\mathbf{g}_o=(1,0,1)$. The other parameters remain the same as those in (a). The spectrum of edge states in (c) shows two counter-propagating Majorana edge states of the helical TSC.}
\label{fig3}
\end{figure*}

Next, we project the pairing Hamiltonian to the new basis, and we find that it also decouples as 
\begin{align}
\mathcal{H}_{\Delta}=\mathcal{H}_{\Delta}^{+}\oplus \mathcal{H}_{\Delta}^{-},
\end{align}
where $\mathcal{H}_{\Delta}^{\pm}$ are given by
\begin{equation}\label{eq-ham-pairing-pm}
\mathcal{H}_{\Delta}^{\pm}= 2\Delta_{\pm}\left\lbrack c_{\pm,\uparrow}^{\dagger}(\mathbf{k}) c_{\pm,\downarrow}^{\dagger}(-\mathbf{k}) - (\uparrow\,\leftrightarrow\,\downarrow) \right\rbrack + \text{h.c.},
\end{equation}
where $\Delta_{\pm}\equiv \Delta_o(\mp id_o^1+d_o^3)$ are the gap strengths in each subspace. In each subspace, it resembles an s-wave superconductor under an effective ``magnetic field'' of the atomic SOC along the out-of-plane direction. It naturally explains the ``zero-field Pauli-limit'' pair-breaking effect of atomic SOC on the orbital-dependent pairings with the $t_{str}\to0$ limit. As a brief conclusion, our results demonstrate that the spin-singlet orbital-dependent pairings occur only in weak atomic SOC electronic systems.

\section{2D Helical superconductivity}
\label{sec-four}
	
In the above sections, the spin-orbit-coupled SCs concerning  inversion symmetry have been comprehensively studied. In addition to that, it will be natural to ask if there exist more interesting superconducting states (e.g.~topological phases) by including an inversion-symmetry breaking to the normal Hamiltonian in Eq.~\eqref{eq-totl-normal-ham}, namely, $\lambda_R\neq0$. For this purpose, in this section, we focus on the Rashba SOC and explore its effect on the spin-orbit-coupled SCs, especially the orbital-dependent pairings. Even though the 2D bulk SC or thin film SC preserves the inversion symmetry, a Rashba SOC appears near an interface between the superconducting layer and the insulating substrate. Remarkably, we find a TRI topological SC (helical TSC) phase generated by the interplay between the two types of SOC (atomic and Rashba) and spin-singlet orbital-dependent pairings. Since TRS is preserved, it belongs to Class DIII according to the ten-fold classification.  On the boundary of the interface, there exists a pair of helical Majorana edge states~\cite{qi_prl_2009,liu_prb_2011,nakosai_prl_2012,deng_prl_2012,zhang_prl_2013,wang_prb_2014,haim_pr_2019,casas_prb_2019,volpez_prr_2020,zhang_prl_2021}.

To explore the topological phases, we consider the normal-state Hamiltonian in Eq.~\eqref{eq-totl-normal-ham}, and the TRI spin-singlet non-unitary pairing symmetry in Eq.~\eqref{eq-delta-TRI-nonunitary-main} for the BdG Hamiltonian~\eqref{eq-bdg-ham}, namely, a real orbital $\mathbf{d}_o$-vector is assumed for the orbital-dependent pairings.

In the $t_{str}\to0$ and $\Delta_s\to0$ limit, the bulk band gap closes only at the $\Gamma$-point for $\mu_c^\pm=\pm\sqrt{\lambda_{soc}^2-4|\Delta_o|^2}$ while no gap-closing happens at other TRI momenta, leading to a topological phase transition. Thus, we conclude that the topological conditions are $\mu_c^-<\mu<\mu_c^+$ and an arbitrary orbital $\mathbf{d}_o$-vector. In Appendix~\ref{app-chern}, we show the $\mathcal{Z}_2$ topological invariant can be analytically mapped to a BdG-version spin Chern number, similar to the spin Chern number in the 2D topological insulators. As mentioned in Sec.~\ref{sec-three} (c), the conservation of $\tau_2$, the U(1) symmetry in the orbital subspace, leads to the decomposition of the BdG Hamiltonian into two blocks for different eigenvalues of $\tau_2$. In each subspace, we can define the BdG Chern number as
\begin{align}
\mathcal{C}_{\pm}=\frac{1}{2\pi}\sum_{\text{filled bands}}\int_{BZ} d\mathbf{k}\cdot \langle\phi_n^{\pm}(\mathbf{k})|i\bm{\nabla}_{\mathbf{k}}|\phi_n^{\pm}(\mathbf{k})\rangle,
\end{align}
with $|\phi_n^{\pm}\rangle$ being the energy eigenstate of $\mathcal{H}_{BdG}^{\pm}$ (see the details in Appendix~\ref{app-chern}). Then the $Z_2$ topological invariant, in this case, is then explicitly given by,
\begin{equation}
\nu\equiv \frac{\mathcal{C}_{+}-\mathcal{C}_{-}}{2},
\end{equation}
where $\mathcal{C}_{\pm}$ are the Chern numbers of the $\pm$ channels. $\nu=1$ corresponds to the TSC phase, shown in Fig.~\ref{fig3} (a). Based on the analysis for the topological condition, we learn that $\Delta_o$ should be smaller than $\lambda_{soc}$. However, as shown in Sec.~\ref{sec-three}, the atomic SOC actually will reduce the $T_c$ of orbital-dependent pairings, which set a guideline to a physically realizable set of parameters, $T_0\gg \lambda_{soc} \gg \Delta_o$, beyond the BCS theory ($\Delta_o\sim 1.76 T_0$). For example, the monolayer FeSe superconductor films on different substrates achieve a very high critical temperature $T_0\sim 70$ K~\cite{ge_natmat_2014}.

As for a more general case with non-zero $\lambda_o$, $t_{str}$ and $\Delta_s$, the BdG Hamiltonian can no longer be decomposed into two decoupled blocks, hence the Chern number approach fails to characterize the $Z_2$ invariant. However, the more general Wilson-loop approach still works (see details in Appendix~\ref{appendix-wilson-loop}). In general, the $Z_2$-type topological invariant of helical superconductivity could be characterized by the Wilson loop spectrum \cite{RuiYU_prb_2011,Benalcazar_prb_2017}, shown in Fig.~\ref{fig3} (b), which demonstrates the non-trivial $Z_2$ index. To verify the helical topological nature, we calculate the edge spectrum in a semi-infinite geometry with $k_y$ being a good quantum number. Fig.~\ref{fig3} (c) confirms clearly that there is a pair of 1D helical Majorana edge modes (MEMs) propagating on the boundary of the 2D system.

\section{TRB non-unitary superconductor}
\label{sec-five}

So far, the TRI non-unitary pairing states are investigated, which exhibit the Pauli-limit violation for in-plane upper critical field and topological phases. Furthermore, in this section, we study the TRB non-unitary pairing states characterized by a complex $\mathbf{d}_o$-vector when both $\Delta_s$ and $\Delta_o$ are real. As it is well known, the experiments by zero-field muon-spin relaxation ($\mu$SR) and the polar Kerr effect (PKE) can provide strong evidence for the observation of spontaneous magnetization or spin polarization in the superconducting states, which indicates a TRB superconducting pairing symmetry. On the theory side, the non-unitary spin-triplet pairing potentials are always adopted to explain the experiments. However, for a spin-singlet SC, a theory with TRB pairing-induced spin-magnetization is in great demand. Addressing this crucial issue is one of the aims of this work, and we find that a spin-singlet TRB non-unitary SCs supports a TRB atomic orbital polarization, which in turn would give rise to spin polarization in the presence of atomic SOC.

\subsection{2D chiral TSC}

We first explore the possible 2D chiral topological phases by considering the simplest case with $\lambda_o=t_{str}=\Delta_s=0$ to demonstrate the essential physics. For the TRB non-unitary pairing, a complex orbital $\mathbf{d}_o$-vector can be generally parameterized as $\mathbf{d}_o=(\cos\theta,0,e^{i\phi}\sin\theta)$. And the relative phase $\phi=\pm\pi/2$ is energetically favored by minimizing the free energy. 

\begin{figure}[t]
\centering
\includegraphics[width=0.9\linewidth]{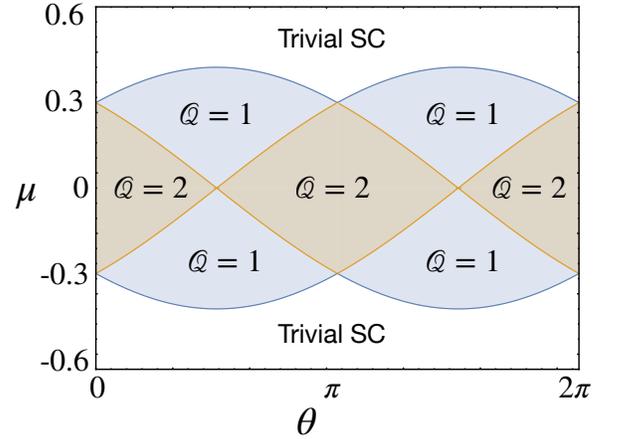}
\caption{Topological chiral superconductivity. We plot the phase diagram in terms of the number of MEMs ($\mathcal{Q}$) of the TSC. Parameters used: $\Delta_o=0.14$, $\lambda_{soc}=0.4$, $\phi=\pm\pi/2$, $\lambda_o=0$, $t_{str}=0$ and $\Delta_s=0$.}
\label{fig4}
\end{figure}

At the $\Gamma$ point, the bulk gap closes at $\mu_{c,i}^\pm=\pm\sqrt{\lambda_{soc}^2-4\vert \Delta_{i}\vert^2}$, where $i=1,2$ and $\Delta_{1,2}=i\Delta_o(\sin\theta\pm\cos\theta)$. Due to TRB, $\mu_{c,1}^\pm\neq \mu_{c,2}^\pm$. Accordingly, we semi-qualitatively map out the phase diagram in Fig.~\ref{fig4} by tuning $\theta$ and $\mu$, and label the different phase regions by the number of Majorana edge modes (MEMs), denoted as $\mathcal{Q}$. When $|\mu|>\text{max}\{|\mu_{c,1}|,|\mu_{c,2}|\}$, the topologically trivial phase is achieved with $\mathcal{Q}=0$. As for $\text{min}\{|\mu_{c,1}|,|\mu_{c,2}|\}<|\mu|<\text{max}\{|\mu_{c,1}|,|\mu_{c,2}|\}$, there is only one MEM on the boundary, corresponding to the $\mathcal{Q}=1$ regions \cite{qi_prb_2010,sau_prl_2010}. When $|\mu|<\text{min}\{|\mu_{c,1}|,|\mu_{c,2}|\}$, there are $\mathcal{Q}=2$ MEMs. The chiral TSC might be detected by anomalous thermal Hall conductivity $K_{xy}=\frac{\mathcal{Q}}{2}\frac{\pi T}{6}$ \cite{meng_prb_2012}.

\subsection{Atomic orbital polarization and spin polarization}

\begin{figure}[t]
\centering
\includegraphics[width=\linewidth]{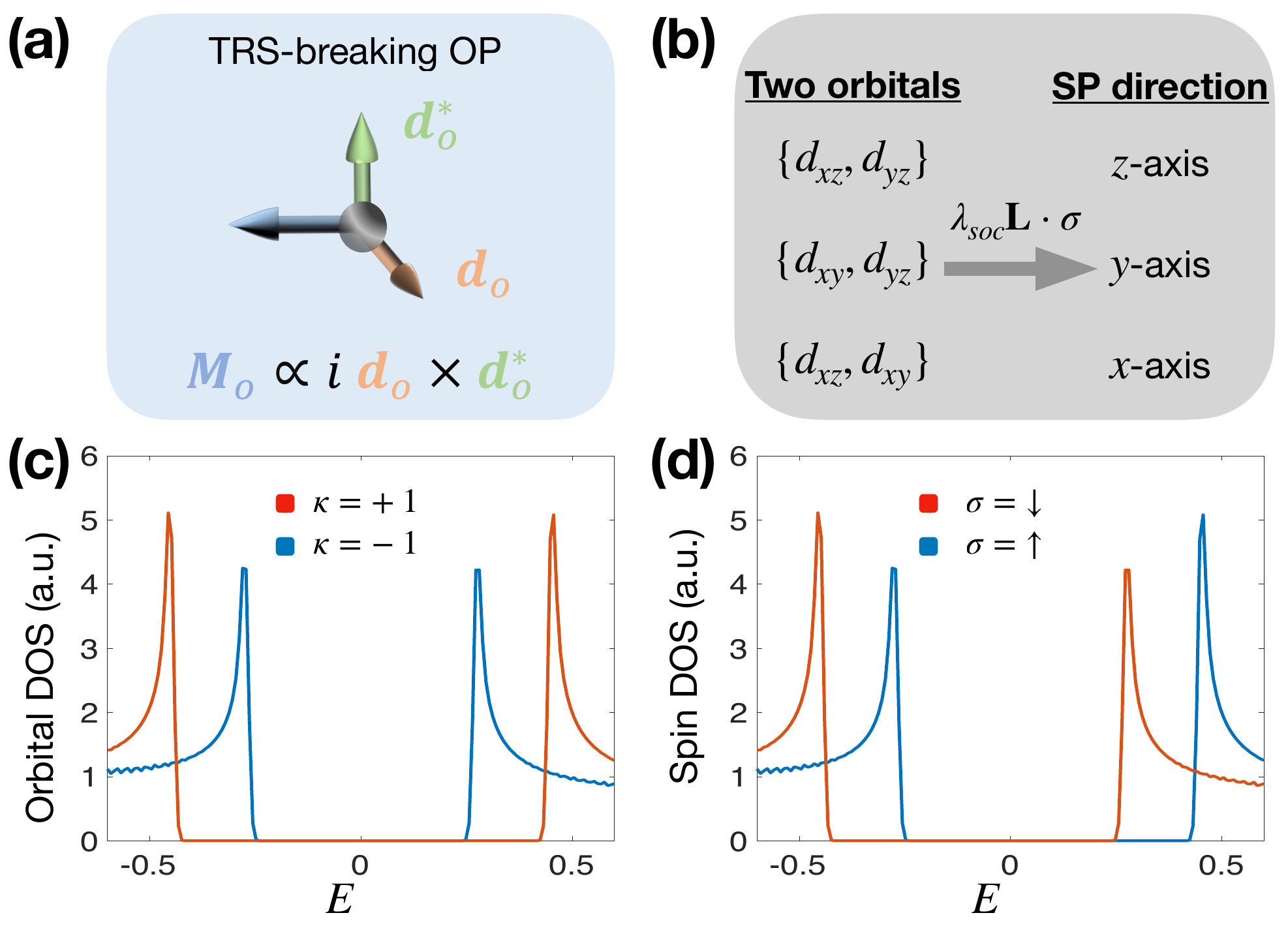}
\caption{(a) Schematic diagram showing the TRB orbital polarization (OP) induced by complex $\mathbf{d}_o$-vector. (b) Spin could be polarized in different directions based on the two active orbitals involved in the pairing. (c) Orbital DOS projected into the chiral $\kappa=\pm 1$ basis, showing a two-gap feature due to TRB. (d) The corresponding spin DOS, shifted relative to the fermi level due to the non-zero effective Zeeman field from the OP. Parameters used: $m=0.5,\mu=-2,\lambda_R=0,\lambda_{soc}=0.2,\Delta_o=0.4,\lambda_o=0, t_{str}=0, \mathbf{d}_o=(1,0,\mathrm{e}^{i\pi/10})$. }
\label{fig5}
\end{figure}

Next, we show how spin-singlet TRB non-unitary pairing can induce spin polarization, and discuss how to identify such pairings by using spin-polarized scanning tunneling microscopy measurements. We assume a TRB complex orbital $\mathbf{d}_o$-vector and find that it can generate the orbital orderings as
\begin{align}\label{eq-AOP-induced-by-d}
\mathbf{M}_o=-i\gamma_1/\alpha_M(\mathbf{d}_o\times \mathbf{d}_o^\ast),
\end{align} 
of which the $y$-component breaks TRS shown in Fig.~\ref{fig5} (a). More precisely, we find that ${M}_o^y \propto \sum_{\mathbf{k},\sigma} \langle \hat{n}_{d_{xz}+id_{yz},\sigma} (\mathbf{k}) - \hat{n}_{d_{xz}-id_{yz},\sigma}(\mathbf{k}) \rangle \neq 0$ indicates the atomic orbital-polarization (OP) (see Appendix~\ref{append::orbital-magnetization} for details). Here, $\hat{n}$ is the density operator of electrons. Once ${M}_o^y$ develops a finite value, it leads to orbital-polarized DOS and two distinct superconducting gaps of the quasi-particle spectrum (Fig.~\ref{fig5} (c), more details below). Therefore, the orbital degree of freedom in spin-singlet SCs plays a similar role as the spin degree of freedom of spin-triplet SCs.

Once the atomic SOC is present, spin-polarization (SP) could be induced indirectly. A possible Ginzburg-Landau term could be
\begin{align}\label{eq:free-energy-spin}
\Delta\mathcal{F}=\alpha_s \vert \mathbf{M}_s\vert^2 + \gamma_{soc} {M}_s^z{M}_o^y ,
\end{align}
with $\alpha_s>0$ and $\gamma_{soc}\neq0$. Here, ${M}_s^z\propto \sum_{\mathbf{k},\tau}\langle \hat{n}_{\tau,\uparrow} - \hat{n}_{\tau,\downarrow} \rangle$. Therefore, the complex orbital $\mathbf{d}_o$-vector can be identified by the spin-resolved density of states (DOS) for spin-singlet superconductors.Minimizing Eq.~\eqref{eq:free-energy-spin}directly leads to $M_s^z = \gamma_{soc}M_o^y/M_s^z$, which indicates the OP-induced spin magnetism. In addition, the direction of SP can be also aligned to $x$ or $y$ axes, discussed later.

To verify the above analysis, we numerically solve the BdG Hamiltonian \eqref{eq-bdg-ham}, $\mathcal{H}_{BdG}\vert E_n(\mathbf{k})\rangle = E_n(\mathbf{k}) \vert E_n(\mathbf{k})\rangle$, where the $n$-th eigenstate is given by $\vert E_n(\mathbf{k})\rangle = (u_{d_{xz},\uparrow}^n,u_{d_{xz},\downarrow}^n,v_{d_{xz},\uparrow}^n,v_{d_{xz},\downarrow}^n,u_{d_{yz},\uparrow}^n,
u_{d_{yz},\downarrow}^n, v_{d_{yz},\uparrow}^n,v_{d_{yz},\downarrow}^n)^T$. Thus, the atomic-orbital and spin-resolved DOS can be calculated as the following,
\begin{equation}
\begin{split}
D_{orbit}^{\kappa}(E) &= \sum_{\sigma,n,\mathbf{k}} \vert u_{\kappa,\sigma}^n\vert^2\delta\left(E- E_n(\mathbf{k})\right),\\
D_{spin}^\sigma(E) &= \sum_{\tau,n,\mathbf{k}} \vert u_{\tau,\sigma}^n \vert^2 \delta\left(E- E_n(\mathbf{k})\right),
\end{split}
\end{equation}
where $u_{\kappa,\sigma}^n =\frac{1}{\sqrt{2}} (u^n_{d_{xz},\sigma}-i \kappa u^n_{d_{yz},\sigma}) $ and $\kappa=\pm1$ for $d_{xz}\pm i d_{yz}$ orbitals. In Fig.~\ref{fig5} (c), $D_{orbit}^{+1}\neq D_{orbit}^{-1}$ indicates that the DOS is orbital-polarized. Remarkably, we also have $D_{spin}^{\uparrow}\neq D_{spin}^{\downarrow}$ due to coupling between electron spin and atomic orbitals, shown in Fig.~\ref{fig5} (d). The difference in orbital DOS acts as an effective Zeeman field for the electron spins, hence shifting the spin DOS relative to the fermi level in opposite directions for up spin and down spin. This interesting phenomenon is quite different from spin-triplet SCs. In TRB spin-triplet SCs, The spin-up channel and spin-down channel will form different symmetric gaps in spin DOS, similar to the two orbital channels in Fig.~\ref{fig5} (c) for our case. Therefore, the spin DOS profiles are distinct in the two cases. As a result, the spin-resolved DOS, which can be probed by spin-resolved STM \cite{wiesendanger2009spin} and muon-spin relaxation \cite{csire2021magneticallytextured,shang2018time}, can serve as a smoking gun evidence to identify TRB due to  complex orbital $\mathbf{d}_o$-vector in multi-orbital SCs.

\section{Discussions and conclusions}
\label{sec-six}

In the end, we briefly discuss the direction of spin-polarization induced by atomic orbital-polarization, summarized in Fig.~\ref{fig5} (b). We consider the three-dimensional subspace of $t_{2g}$ orbitals spanned by $\{d_{yz},d_{xz},d_{xy}\}$, where the matrix form of the angular momentum operators $\mathbf{L}$ reads~\cite{lee_prb_2010},
\begin{align}
L_x = \begin{pmatrix}
0 & 0  & 0 \\ 
0 & 0  & i  \\ 
0 & -i & 0 
\end{pmatrix}, \;
L_y = \begin{pmatrix}
0 & 0  & -i \\ 
0 & 0  & 0  \\ 
i & 0 & 0 
\end{pmatrix},  \;
L_z = \begin{pmatrix}
0  & i  & 0 \\ 
-i & 0  & 0  \\ 
0  & 0  & 0 
\end{pmatrix},  
\end{align}
which satisfy the commutation relation $[L_m,L_n]=-i\epsilon_{mnl}L_l$. Therefore, the spin-orbit coupling for a system with the $t_{2g}$ orbitals is given by,
\begin{align}
H_{soc} = \lambda_{soc} \mathbf{L}\cdot\bm{\sigma}.
\end{align}
Then, let us consider a two-orbital system, the above SOC Hamiltonian will be reduced to,
\begin{align}
\begin{cases}
\text{For }\{d_{yz},d_{xz}\}:\; H_{soc} = -\lambda_{soc}\tau_2 \sigma_3, \\
\text{For }\{d_{yz},d_{xy}\}:\; H_{soc} = \lambda_{soc} \tau_2 \sigma_2, \\
\text{For }\{d_{xz},d_{xy}\}:\; H_{soc} = -\lambda_{soc} \tau_2 \sigma_1.
\end{cases}
\end{align}
Therefore, in the above three cases, the spin-polarization is pointed to $z,y,x$-axis, respectively. Because the atomic orbital polarization is induced by the complex orbital $\mathbf{d}_o$-vector as $(0,M_o^y,0)\propto i\mathbf{d}_o^\ast \times \mathbf{d}_o$.

To summarize, we establish a phenomenological theory for spin-singlet two-band SCs and discuss the distinct features of both TRI non-unitary pairings and TRB non-unitary pairings by studying the effects of atomic spin-orbit coupling (SOC), lattice strain effect, and Rashba SOC. Practically, we demonstrate that the stability of orbital-dependent pairing states could give birth to the non-unitary pairing states in a purely spin-singlet SC. Remarkably, the interplay between atomic SOC and orbital-dependent pairings is also investigated and we find a new spin-orbit-coupled SC with spin-singlet non-unitary pairing. For this exotic state, there are mainly three features. 
Firstly, the atomic SOC could enlarge the in-plane upper critical field compared to the Pauli limit. A new effect dubbed as ``zero-field Pauli limit'' for orbital-dependent pairings is discovered. 
Secondly, topological chiral or helical superconductivity could be realized even in the absence of external magnetic fields or Zeeman fields.
Furthermore, a spontaneous TRB SC could even generate a spin-polarized superconducting state that can be detected by  measuring the spin-resolved density of states. 
We hope our theory leads to a deeper understanding of spin-singlet non-unitary SCs.

Our theory might have potential applications to the intriguing Sr$_2$SuO$_4$ \cite{luke1998,xia2006}, LaNiGa$_2$ \cite{weng2016}, iron-based SCs \cite{zaki2019,grinenko2020} and ultra-cold atomic systems with large spin alkali and alkaline-earth fermions \cite{ho_prl_1999,wu_prl_2003,desalvo_prl_2010,taie_prl_2010,gorshkov_np_2010}.

\section{Acknowledgments}

We thank J.-L.~Lado, R.-X.~Zhang and C.-X.~Liu for helpful discussions. We especially acknowledge J.-L.~Lado’s careful reading of the manuscript. D.-H.X. was supported by the NSFC (under Grant Nos. 12074108 and 11704106).

\bibliographystyle{apsrev4-2}
\bibliography{ref}

\clearpage
\appendix

\section{Toy model for two-band superconducting phase diagrams}
In this part of the appendix, we explore a possible superconducting phase diagram including the non-unitary pairing states in the GL framework. 
Here we assume a two-band SC with
\begin{align}\label{eq-delta-TRI-nonunitary}
\Delta_{tot}=\left\lbrack \Delta_s\tau_0+\Delta_o(d_o^1\tau_1+d_o^3\tau_3)\right\rbrack (i\sigma_2).
\end{align}
In terms of the superconducting order parameters $\{\Delta_s,\Delta_o,\mathbf{d}_o=(d_o^1,0,d_o^3)\}$ and the order parameter for the orbital orderings $\mathbf{M}_o \propto \sum_{\mathbf{k},\sigma} \langle c^\dagger_{a\sigma}(\mathbf{k}) \bm{\tau}_{ab} c_{b\sigma }(\mathbf{k}) \rangle$, the total GL free energy can be constructed to address the homogeneous superconducting phase without external magnetic fields,
\begin{align}
\mathcal{F}[\Delta_s,\Delta_o,\mathbf{d}_o,\mathbf{M}_o] = \mathcal{F}_0 +\mathcal{F}_b + \mathcal{F}_{o} ,
\end{align}
where
\begin{equation}
\begin{split}
\mathcal{F}_0 &=\frac{1}{2}\alpha(T) \vert \Delta_o\vert^2 + \frac{1}{2}\alpha'(T)\vert \Delta_s\vert^2 + \frac{1}{2}\alpha_M|\mathbf{M}_o|^2 \\
& + \frac{1}{4}\beta\vert\Delta_o\vert^4 +\frac{1}{4}\beta'\vert\Delta_s\vert^4 +\beta'' |\Delta_s|^2|\Delta_o|^2 \\
& + \beta_o|d_o^1|^4+\beta_o'|d_o^3|^4,
\end{split}
\end{equation}
where $\vert\mathbf{d}_o\vert=1$ is adopted, $\alpha(T)=\alpha_0(T/T_{c1}-1)$, $\alpha'(T)=\alpha'_0(T/T_{c2}-1)$ and the coefficients $\alpha_0$, $\alpha'_0$, $\alpha_M$, $\beta$, $\beta'$, $\beta''$, $\beta_o$, $\beta_o'$ are all positive. $T_{c1},T_{c2}$ are critical temperatures in orbital-dependent and orbital-independent channels respectively, which are in general different from each other. And $\alpha_M>0$ means that there is no spontaneous atomic orbital polarization. In the superconducting state with both non-zero $\Delta_s$ and $\Delta_o$ developed already, additionally, there are two possible ways to pursue the spontaneous TRB, denoted as $\mathcal{F}_{b}$ and $\mathcal{F}_{o}$. Firstly, we consider the $\mathcal{F}_{b}$ term
\begin{align}
\label{eq-gl-free-energy-f3}
\mathcal{F}_{b} &=b_1\Delta_s^\ast\Delta_o + b_2(\Delta_s^\ast \Delta_o)^2  + \text{h.c.},
\end{align}
where the sign of $b_2$ determines the breaking of TRS. Here we focus on the generic case where $\Delta_s$ and $\Delta_o$ belong to different symmetry representations so that there is no linear order coupling between them, i.e. $b_1=0$. Given $b_1=0$ and $b_2>0$, we have a $\theta_{o}=\pm\pi/2$ relative phase difference between $\Delta_s$ and $\Delta_o e^{i\theta_{o}}$~\cite{wang_prl_2017}, which gives to the achievement of the TRB unitary pairing state ($\Delta_s\in\mathds{R}, \Delta_o\sim i, \mathbf{d}_o\in \mathds{R}$).

More generally, a TRB non-unitary SC arises from the non-zero bilinear $b_1$-term, which is symmetry-allowed only when $\Delta_s$ and $\Delta_o$ belong to the same symmetry representation of the crystalline symmetry group. Namely, the case with $b_1\neq0$ and $b_2>0$ can pin the phase difference $\theta_{o}$ to an arbitrary nonzero value, i.e., $\theta_{o}\in (0,\pi)$. Then, this case can also give rise to TRB non-unitary pairing with ($\Delta_s\in\mathds{R},\Delta_t\in\mathds{C},\mathbf{d}_o\in \mathds{R}$) or ($\Delta_s\in\mathds{R},\Delta_t\in\mathds{C},\mathbf{d}_o\in \mathds{C}$). On the other hand, the $b_2<0$ situation makes TRI non-unitary pairing states ($\Delta_s\in\mathds{R}, \Delta_o\in \mathds{R}, \mathbf{d}_o\in \mathds{R}$).

 However, even in the case with $b_2<0$, we still have an alternative approach to reach TRB pairing states, driven by the
$\mathcal{F}_{o}$ term
\begin{align}
\label{eq-gl-free-energy-f4}
\mathcal{F}_{o} &= \gamma_0 \vert \mathbf{d}_o\times \mathbf{d}_o^\ast \vert ^2 + i\gamma_1\mathbf{M}_o\cdot(\mathbf{d}_o\times \mathbf{d}_o^\ast) + \text{h.c.},
\end{align}
where the sign of $\gamma_0$ identifies the TRB due to a complex $\mathbf{d}_o$. In particular $\gamma_0<0$ results in a TRB non-unitary state ($\Delta_s\in \mathds{R}, \Delta_o\in\mathds{R}, \mathbf{d}_o\in\mathds{C}$).

\begin{figure}[h]
\centering
\includegraphics[width=\linewidth]{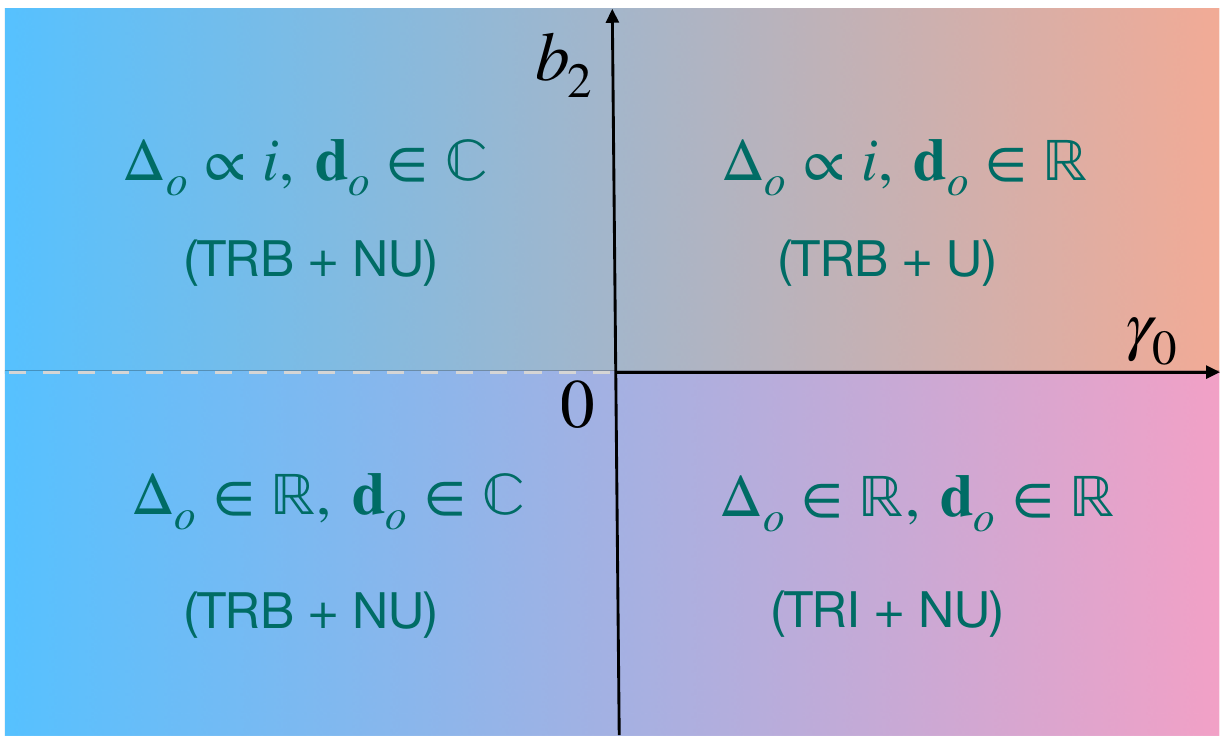}
\caption{Schematic superconducting phase diagrams on the $b_2$-$\gamma_0$ plane when $b_1=0$ and $\Delta_s$ is real and non-zero. Here, TRB and TRI are short for TR-breaking and TR-invariant, respectively; U and NU represent unitary and non-unitary, respectively.
}
\label{fig:my_label}
\end{figure}

We summarize many of the possible interesting superconducting phases in Fig.~\ref{fig:my_label},  which schematically shows a superconducting phase diagram as a function of $b_2$ and $\gamma_0$ by setting $b_1=0$, i.e. the generic case where $\Delta_o,\Delta_s$ belong to different representations. Notice that this phase diagram characterized by $b_2$ and $\gamma_0$ does not contain the TRI unitary pairing phase.

\section{Derivation of $T_c$ from linearized gap equation}
\label{append-Tc-derivation}

\begin{widetext}
Starting from the generic Hamiltonian, containing atomic SOC, generic $\tilde{\bf g}=(\tilde{g}_1,0,\tilde{g}_2)$ with $|\tilde{\bf g}|=1$ and in-plane magnetic field,
\begin{equation}	H_0(\mathbf{k})=\epsilon(\mathbf{k})+\lambda_{soc}\sigma_3\tau_2+\lambda(\tilde{g}_1\tau_1+\tilde{g}_3\tau_3)+h\sigma_1.
\end{equation}
The Matsubara Green's function for electrons is
\begin{equation}
\begin{split}
G_e(\mathbf{k},i\omega_n)&=[i\omega_n-\mathcal{H}_0(\mathbf{k})]^{-1}\\
&=\frac{P_{---}}{i\omega_n-\epsilon_{\mathbf{k}}+E_-}
+\frac{P_{+-+}}{i\omega_n-\epsilon_{\mathbf{k}}+E_+}
+\frac{P_{-+-}}{i\omega_n-\epsilon_{\mathbf{k}}-E_-}
+\frac{P_{+++}}{i\omega_n-\epsilon_{\mathbf{k}}-E_+},
\end{split}
\end{equation}
where the projection operator 
\begin{equation}
P_{\alpha\beta\gamma}=\frac{1}{4}[1+\alpha(\tilde{g}_1\sigma_1\tau_1+\tilde{g}_3\sigma_1\tau_3)]\cdot[1+\frac{\beta}{E_\gamma}(\lambda_{soc}\sigma_3\tau_2+\lambda_o(\tilde{g}_1\tau_1+\tilde{g}_3\tau_3)+h\sigma_1)],
\end{equation}
with $\alpha,\beta,\gamma \in\{+,-\}$ and $E_\gamma=\sqrt{\lambda_{soc}^2+(\lambda+\gamma h)^2}$. The Green's function for hole is $G_h(\mathbf{k},i\omega_n)=-G_e^\ast(\mathbf{k},i\omega_n)$. Here $\omega_n=(2n+1)\pi k_BT$.

The linearized gap equation is given by
\begin{equation}
\Delta_{s_1,s_2}^{a,b}(\mathbf{k}) = -\frac{1}{\beta} \sum_{\omega_n}\sum_{s_1'a',s_2'b'} V^{s_1a,s_2b}_{s_1'a',s_2'b'}(\mathbf{k},\mathbf{k}')
\times 
\left\lbrack G_e(\mathbf{k}',i\omega_n) \Delta(\mathbf{k}') G_h(-\mathbf{k}',i\omega_n) \right\rbrack_{s_1'a',s_2'b'},
\end{equation}
where the generic attractive interaction can be expanded as
\begin{align}
V^{s_1a,s_2b}_{s_1'a',s_2'b'}(\mathbf{k},\mathbf{k}') = -v_0\sum_{\Gamma,m} \lbrack \mathbf{d}_o^{\Gamma,m}(\mathbf{k})\cdot\bm{\tau}i\sigma_2 \rbrack_{s_1a,s_2b} \lbrack \mathbf{d}_o^{\Gamma,m}(\mathbf{k}')\cdot\bm{\tau}i\sigma_2 \rbrack_{s_1'a',s_2'b'},
\end{align}
where $v_0>0$ and $\Gamma$ labels the irreducible representation with $m$-dimension of crystalline groups. The linearized gap equation is reduced to $v_0\chi(T)-1 = 0$ where $\chi(T)$ is the superconductivity susceptibility. We have
\begin{itemize}
\item For orbital-independent pairing:
\begin{align}
\chi(T)_s = -\frac{1}{\beta}\sum_{\mathbf{k},\omega_n} \text{Tr}\left\lbrack (\psi_s(\mathbf{k})i\sigma_2)^\dagger G_e(\mathbf{k},i\omega_n) (\psi_s(\mathbf{k})i\sigma_2) G_h(-\mathbf{k},i\omega_n) \right\rbrack.
\end{align}
\item For orbital-dependent pairing:
\begin{align}
\chi(T)_o = -\frac{1}{\beta}\sum_{\mathbf{k},\omega_n} \text{Tr}\left\lbrack (\mathbf{d}_o(\mathbf{k})\cdot\bm{\tau}i\sigma_2)^\dagger G_e(\mathbf{k},i\omega_n) (\mathbf{d}_o(\mathbf{k})\cdot\bm{\tau}i\sigma_2) G_h(-\mathbf{k},i\omega_n) \right\rbrack.
\end{align}
\end{itemize}

Then we take the standard replacement, 
\begin{align}
\sum_{\mathbf{k},\omega_n} \to \frac{N_0}{4}\int_{-\omega_D}^{+\omega_D} d\epsilon \iint_{S}d\Omega \, \sum_{\omega_n},
\end{align}
where $N_0$ is the density of states at Fermi surface, $\Omega$ is the solid angle of $\mathbf{k}$ on Fermi surfaces and $\omega_D$ the Deybe frequency. We will also be making use of,
\begin{align}
&-\frac{N_0}{\beta} \int_{-\omega_D}^{+\omega_D} \sum_{\omega_n} d\epsilon G_e^+(\mathbf{k},i\omega_n) G_h^+(\mathbf{k},i\omega_n) = -\frac{N_0}{\beta} \int_{-\omega_D}^{+\omega_D} \sum_{\omega_n} G_e^-(\mathbf{k},i\omega_n) G_h^-(\mathbf{k},i\omega_n) = \chi_0(T),\\
&-\frac{N_0}{\beta} \int_{-\omega_D}^{+\omega_D} \sum_{\omega_n} d\epsilon G_e^-(\mathbf{k},i\omega_n) G_h^+(\mathbf{k},i\omega_n) = -\frac{N_0}{\beta} \int_{-\omega_D}^{+\omega_D} \sum_{\omega_n} G_e^+(\mathbf{k},i\omega_n) G_h^-(\mathbf{k},i\omega_n) = \chi_0(T) + N_0\mathcal{C}_0(T),
\end{align}
where $\chi_0(T)=N_0\ln\left( \tfrac{2e^\gamma \omega_D}{\pi k_B T} \right)$, $\gamma=0.57721\cdots$ the Euler-Mascheroni constant  and $\mathcal{C}_0(T)=\text{Re}\lbrack \psi^{(0)}(\tfrac{1}{2}) - \psi^{(0)}(\tfrac{1}{2}+i\tfrac{E(\mathbf{k})}{2\pi k_BT}) \rbrack$ with $\psi^{(0)}(z)$ being the digamma function.

For orbital-independent pairing considered in the main text $\Delta_s\tau_0i\sigma_2$, we have 
\begin{equation}
\begin{split}
\chi_s(T)=\chi_0(T)&+\frac{N_0}{2}\left[\mathcal{C}_0\left(T,\frac{E_+-E_-}{2}\right)+\mathcal{C}_0\left(T,\frac{E_++E_-}{2}\right)\right]\\
& +\frac{N_0}{2}\left[\mathcal{C}_0\left(T,\frac{E_+-E_-}{2}\right)-\mathcal{C}_0\left(T,\frac{E_++E_-}{2}\right)\right]\times\frac{\lambda^2+\lambda_{soc}^2-h^2}{E_+E_-}\\
\equiv\chi_0(T)&+N_0f_s(T,\lambda_{soc},\lambda,h).
\end{split}
\end{equation}
In order to look at the effect of $\lambda$ on the Pauli limit, we could Taylor expand $f_s(T,\lambda_{soc},\lambda,h)$ for small $\lambda$:
\begin{equation}		f_s(T,\lambda_{soc},\lambda,h)=f_s(T,\lambda_{soc},0,h)+F(T,\lambda_{soc},h)\lambda^2+\mathcal{O}(\lambda^4), 
\end{equation}
with 
\begin{equation}
\begin{split}
F(T,\lambda_{soc},h)&=\psi^{(2)}(\tfrac{1}{2})\frac{\lambda_{soc}^2h^2}{4\pi k_B^2T^2(\lambda_{soc}^2+h^2)^2}\\
&-\mathrm{Re}\{\psi^{(0)}(\tfrac{1}{2})-\psi^{(0)}(\tfrac{1}{2}+i\tfrac{\sqrt{\lambda_{soc}^2+h^2}}{2\pi k_BT})\}\frac{4\lambda_{soc}^2h^2}{(\lambda_{soc}^2+h^2)^3}\\
&+\mathrm{Im}\{\psi^{(1)}(\tfrac{1}{2}+i\tfrac{\sqrt{\lambda_{soc}^2+h^2}}{2\pi k_BT})\}\frac{\lambda_{soc}^2h^2}{2\pi k_BT(\lambda_{soc}^2+h^2)^{5/2}}.
\end{split}
\end{equation}
This is used in the main text.

For orbital-dependent pairing $\Delta_o(d_1\tau_1+d_3\tau_3)i\sigma_2$ with $\mathbf{d}_o=\tilde{\bf g}$, we have 
\begin{equation}
\begin{split}
\chi_o(T)=\chi_0(T)&+\frac{N_0}{2}\left[\mathcal{C}_0\left(T,\frac{E_+-E_-}{2}\right)+\mathcal{C}_0\left(T,\frac{E_++E_-}{2}\right)\right]\\
& +\frac{N_0}{2}\left[\mathcal{C}_0\left(T,\frac{E_+-E_-}{2}\right)-\mathcal{C}_0\left(T,\frac{E_++E_-}{2}\right)\right]\times\frac{\lambda^2-\lambda_{soc}^2-h^2}{E_+E_-}\\
\equiv\chi_0(T)&+N_0f_o(T,\lambda_{soc},\lambda,h).
\end{split}
\end{equation}

\end{widetext}

\section{Strain effect on $T_c$ and pairing symmetry}
 \label{app-strain-effect}

\begin{figure*}[!htbp]
\centering
\includegraphics[width=0.75\linewidth]{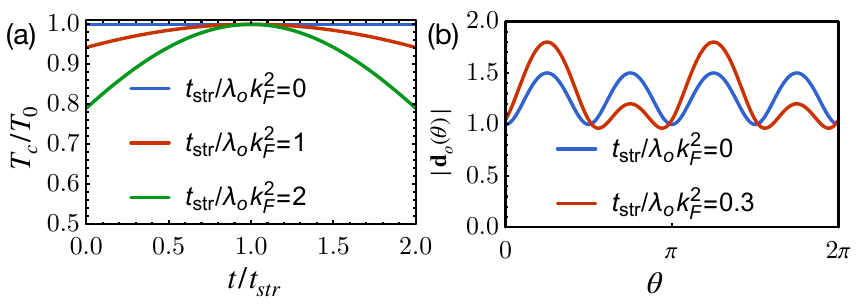}
\caption{
(a) shows the suppression of $T_c$ for different strain strengths. Here $\mathbf{d}_o=\mathbf{g}_o+\frac{t}{\lambda_o}\mathbf{g}_{str}$ whereas $\tilde{\bf g}=\mathbf{g}_o+\frac{t_{str}}{\lambda_o}\mathbf{g}_{str}$.  
(b) shows the symmetry breaking of the SC gap from $C_4$ to $C_2$ due to the existence of the external strain. 
We have chosen $\mathbf{g}_o=(3k_xk_y, 0, k_x^2-k_y^2)$ and the strain parameter $\phi=0$ in $\mathbf{g}_{str}$.
}
\label{fig-append-C}
\end{figure*}

 The strain effect characterized by Eq.~(\ref{eq-strain}) in the main text can be absorbed into the orbital hybridization vector $\mathbf{g}_o$ and gives rise to an effective $\tilde{\bf g}\equiv \mathbf{g}_o+t_{str}/\lambda_o(\sin{2\phi},0,\cos{2\phi})$. Then in the absence of SOC terms, the corrected critical temperature $T_c$ due to the strain and hybridization effects is perturbatively given by 
\begin{align}    
\ln{\left(\frac{T_c}{T_0}\right)}=\int \int_S d\Omega\ \mathcal{C}_0(T_0)\left(|\mathbf{d}_o|^2-|\mathbf{d}_o\cdot\hat{\tilde{\bf g}}|^2\right),
\end{align}
where $T_0$ is the critical temperature without strain or hybridization and the integration is over the solid angle of $\mathbf{k}$ over the Fermi surface. Similar to previous discussions, the strain generally suppresses the critical temperature when $\tilde{\bf g}$ is not exactly parallel to $\mathbf{d}_o$, as shown in Fig.~\ref{fig-append-C} (a). For non-zero strain, the $T_c$ is not suppressed when $\mathbf{d}_o||\tilde{\bf g}$. Fig.~\ref{fig-append-C} (b) shows the symmetry breaking pattern of the $|\mathbf{d}_o|$, which is proportional to the SC gap (the proportionality constant has been normalized to 1 in the figure), around the Fermi surface. The strain would reduce the symmetry from $C_4$ to $C_2$, as expected.

\section{TSC with $\Delta_s=0,\lambda_o=0$}
\label{app-chern}

To demonstrate the topology, we also show a simple case with $\Delta_s=0$ and $\lambda_o=0$, where the $Z_2$ can be characterized analytically.

In this section, we focus on the simplified case without orbital independent pairing or orbital hybridization. In Fig.~\ref{fig3} (c), we calculate the edge spectrum with $k_x$ being a good quantum number in a semi-infinite geometry, and it shows the corresponding bulk band structure together with two counter-propagating MEMs. The bulk topology of the 2D helical TSC phase is characterized by the $Z_2$ topological invariant $\nu$, which can be extracted by calculating the Wilson-loop spectrum. And, $\nu=1$ mod $2$ characterizes the helical TSC. In Fig.~\ref{fig3} (b), we plot the evolution of $\theta$ as a function of $k_y$, and the winding pattern indicates the topological $Z_2$ invariant $\nu=1$.

On the other hand, with $\Delta_s=0$, which is the case if we only consider on-site attractive interactions between electrons \cite{liu_prl_2017,hu_cp_2019}, the BdG Hamiltonian~\eqref{eq-bdg-ham} can be decomposed into two orbital subspaces that are related through time-reversal transformation. Each of these blocks has a well-defined Chern number because each block alone breaks TRS. The two Chern numbers can then be used to define the $Z_2$ invariant of the whole BdG system. The detailed procedures are the following.

For the normal Hamiltonian given in Eq.~\eqref{Eq::normal_H}, we have $[\mathcal{H}_0,\tau_2]=0$. As a result, we can project the normal Hamiltonian $\mathcal{H}_0$ in Eq.~\eqref{Eq::normal_H} into block-diagonal form corresponding to the $\pm 1$ eigenvalues of $\tau_2$ by using the basis transformation $\mathcal{U}=\sigma_0\otimes \frac{1}{\sqrt{2}}\begin{bmatrix}1&-i\\1&i \end{bmatrix}$. 
The new basis is given by 
\begin{align}
\Tilde{\Psi}^\dagger(\mathbf{k})=(c_{+,\uparrow}^\dagger,c_{+,\downarrow}^\dagger,c_{-,\downarrow}^\dagger,c_{-,\uparrow}^\dagger),
\end{align}
where $c_{\pm,s}^\dagger\equiv\frac{1}{\sqrt{2}}(c_{d_{xz},s}^\dagger\mp i c_{d_{yz},s}^\dagger)$. 
On this basis, the normal Hamiltonian is given by 
\begin{align}
\mathcal{H}_0=\mathcal{H}_0^{+}\oplus \mathcal{H}_0^{-},
\end{align} 
where $\mathcal{H}_0^{\pm}$ are given by
\begin{align}
\mathcal{H}_0^{\pm}=\epsilon(\mathbf{k}) + \lambda_R(k_x\sigma_2-k_y\sigma_1) \mp \lambda_{soc}\sigma_3.
\end{align}
Note that the time-reversal transforms $\mathcal{H}_0^{\pm}(\mathbf{k})$ to $\mathcal{H}_0^{\mp}(-\mathbf{k})$. In the new basis the pairing Hamiltonian also decouples as $\mathcal{H}_{\Delta}=\mathcal{H}_{\Delta}^{+}\oplus \mathcal{H}_{\Delta}^{-}$ with $\mathcal{H}_{\Delta}^{\pm}$ given by
\begin{equation}\label{eq-ham-pairing-pm}
\mathcal{H}_{\Delta}^{\pm}= 2\Delta_{\pm}\left\lbrack c_{\pm,\uparrow}^{\dagger}(\mathbf{k}) c_{\pm,\downarrow}^{\dagger}(-\mathbf{k}) - (\uparrow\,\leftrightarrow\,\downarrow) \right\rbrack + \text{h.c.},
\end{equation}
where $\Delta_{\pm}\equiv \Delta_o(\mp id_o^1+d_o^3)$ are the gap strengths in each subspace. Therefore, the Bogoliubov de-Gennes (BDG) Hamiltonian takes the following block-diagonal form, 
\begin{align}\label{eq-ham-bdg-ambasis}
\mathcal{H}_{BdG} = \mathcal{H}_{BdG}^{+} \oplus \mathcal{H}_{BdG}^{-},
\end{align}
where 
\begin{align}
\mathcal{H}_{BdG}^{\pm}(\mathbf{k})&=(\epsilon(\mathbf{k})\mp\lambda_{soc} \sigma_3)\gamma_3 + \lambda_R (k_x\sigma_2\gamma_3 
-k_y\sigma_1\gamma_0) \nonumber \\
& \pm 2 d_1\sigma_2\gamma_1-2d_3\sigma_2\gamma_2,
\end{align}
with $\gamma_\mu$ being the  Pauli matrices in the particle-hole space. The Nambu basis is  $\Psi_{\pm}^\dagger(\mathbf{k})=(c_{\pm,\uparrow}^\dagger(\mathbf{k}),c_{\pm,\downarrow}^\dagger(\mathbf{k}),c_{\pm,\uparrow}(-\mathbf{k}),c_{\pm,\downarrow}(-\mathbf{k}))$. Each subspace has its own particle-hole symmetry.

By symmetry, the 2D BdG Hamiltonian in Eq.~\eqref{eq-ham-bdg-ambasis} belongs to Class DIII of the A-Z classification\cite{ryu_jip_2010,chiu_rmp_2016} for topological insulators and superconductors because both TRS and particle-hole symmetry are preserved. However, it is not the case for our model. The BdG Hamiltonian here could exhibit topological states with $Z_2$ type topological invariant, which can be defined as the following. In each subspace, we define the BdG Chern number as
\begin{align}
\mathcal{C}_{\pm}=\frac{1}{2\pi}\sum_{\text{filled bands}}\int_{BZ} d\mathbf{k}\cdot \langle\phi_n^{\pm}(\mathbf{k})|i\bm{\nabla}_{\mathbf{k}}|\phi_n^{\pm}(\mathbf{k})\rangle,
\end{align}
with $|\phi_n^{\pm}\rangle$ being the energy eigenstate of $\mathcal{H}_{BdG}^{\pm}$. Then the $Z_2$ invariant, in this case, is then explicitly given by,
\begin{equation}
\nu\equiv \frac{\mathcal{C}_{+}-\mathcal{C}_{-}}{2},
\end{equation}
where $\mathcal{C}_{\pm}$ are the Chern numbers of the $\pm$ channels. This has been discussed in the main text.

\section{Wilson loop calculation for $\mathcal{Z}_2$ TSC}
\label{appendix-wilson-loop}

In the thermodynamics limit, the Wilson loop operator along a closed path $p$ is expressed as
\begin{align}
\mathcal{W}_{p}  = \mathcal{P}\exp\left[ i {\oint_{p} } \mathcal{A} (\mathbf{k} ) \,d \mathbf{k} \right] ,
\end{align}
where $\mathcal{P}$ means path ordering and $\mathcal{A} (\mathbf{k} )$ is the non-Abelian Berry connection
\begin{align}
\mathcal{A} ^{nm} (\mathbf{k}) = i \langle \phi ^{n} (\mathbf{k}) | \nabla _{\mathbf{k} } |\phi ^{m} (\mathbf{k}) \rangle,
\end{align}
with $|\phi ^{m,n} (\mathbf{k}) \rangle$ the occupied eigenstates. The Wilson line element is defined as 
\begin{align}
G ^{nm} (\mathbf{k}) = \langle \phi ^{n} (\mathbf{k}+\Delta\mathbf{k}) |  \phi ^{m} (\mathbf{k}) \rangle  ,
\end{align}
where the $ \mathbf{k} = ( k_{x},k_{y} ) $, and $ \Delta \mathbf{k} =(0, 2\pi / N_{y})  $ is the steps. In the discrete case, the Wilson loop operator on a path along $k_y$ from the initial point $\mathbf{k}$ to the final point  $\mathbf{k}+(0,2\pi)$ can be written as $\mathcal{W} _{y,\mathbf{k}} =  G (\mathbf{k}+(N_y-1)\Delta\mathbf{k} ) G (\mathbf{k}+(N_y-2)\Delta\mathbf{k} )...  G (\mathbf{k}+\Delta\mathbf{k} ) G (\mathbf{k})$, which satisfies the eigenvalue equation
\begin{equation}
\mathcal{W} _{y,\mathbf{k}} | \nu^{j}_{y,\mathbf{k}} \rangle =e^{ i 2 \pi \nu^{j}_{y}(k_x) } |\nu^{j}_{y,\mathbf{k}} \rangle 
\end{equation}
The phase of eigenvalue $ \theta = 2 \pi \nu^{j}_{y}(k_x) $ is the Wannier function center.

\section{Spin and orbital magnetizations: $\mathbf{M}_s$ and $\mathbf{M}_o$}
\label{append::orbital-magnetization}

In this section, we show the definition of spin and orbital magnetization at the mean-field level. The spin magnetization in orbital-inactive systems takes the form 
\begin{align}
\mathbf{M}_s\propto \sum_{\mathbf{k},s_1,s_2}\langle c_{s_1}^{\dagger}(\mathbf{k})\bm{\sigma}_{s_1s_2}c_{s_2}(\mathbf{k})\rangle,
\end{align}
which tells us the magnetic moments generated by spin polarization. Similarly, the orbital magnetization in orbital-active system is given by
\begin{align}
\mathbf{M}_o\propto \sum_{\mathbf{k},s,a,b}\langle c_{s,a}^{\dagger}(\mathbf{k})\bm{\tau}_{ab}c_{s,b}(\mathbf{k})\rangle.
\end{align} 
The different components of the orbital magnetization vector represent different orders in the SC ground state. More specifically, we have 
\begin{align}
M_o^x&=\sum_{\mathbf{k},s}\langle c_{s,d_{xz}}^{\dagger}c_{s,d_{yz}}+c_{s,d_{yz}}^{\dagger}c_{s,d_{xz}}\rangle, \\ M_o^y&=-i\sum_{\mathbf{k},s}\langle c_{s,d_{xz}}^{\dagger}c_{s,d_{yz}}-c_{s,d_{yz}}^{\dagger}c_{s,d_{xz}}\rangle \\
&=\frac{1}{2}\sum_{\mathbf{k},s}\langle \hat{n}_{s,d_{xz}+id_{yz}}-\hat{n}_{s,d_{xz}-id_{yz}}\rangle, \\
M_o^z&=\sum_{\mathbf{k},s}\langle c_{s,d_{xz}}^{\dagger}c_{s,d_{xz}}-c_{s,d_{yz}}^{\dagger}c_{s,d_{yz}}\rangle. 
\end{align}
We see that $M_o^{x,z}$ breaks the $C_4$ rotation symmetry and $M_o^y$ breaks TRS. In our work, we only consider the possibility of spontaneous TRS breaking, thus the $M_o^{x,z}$ will not couple to the superconducting order parameters, which are required to be invariant under $C_n$. Because $M_o^y$ breaks TRS so that it could be coupled to the superconducting order parameters, which spontaneously breaks TRS. This is one of the main results of our work,
\begin{align}
(0,M_o^y,0)\propto i\mathbf{d}_o^\ast \times \mathbf{d}_o,
\end{align}
where the complex orbital $\mathbf{d}_o$-vector breaks TRS.

\end{document}